\documentclass[journal,twoside]{IEEEtran}

\ifCLASSINFOpdf
\usepackage[pdftex]{graphicx}

\else

\fi

\usepackage[cmex10]{amsmath}
\usepackage{amssymb}   
\usepackage{amsxtra}
\usepackage{amscd}
\usepackage{amsthm}
\usepackage{textcomp}
\usepackage{graphicx}

\usepackage{array}  

\usepackage{multirow}  

\usepackage{amsmath}
\usepackage{cite}

\usepackage{url}

\usepackage{color,soul} 
\soulregister\cite7
\soulregister\ref7
\soulregister\pageref7

\begin{document}

\title{{\huge Wireless Communications Through Reconfigurable \\Intelligent Surfaces}}

\author{Ertugrul Basar, \IEEEmembership{Senior Member, IEEE},
Marco Di Renzo, \IEEEmembership{Senior Member, IEEE},
Julien de Rosny, 
Merouane Debbah, \IEEEmembership{Fellow, IEEE},
Mohamed-Slim Alouini, \IEEEmembership{Fellow, IEEE}, and
Rui Zhang, \IEEEmembership{Fellow, IEEE}

\thanks{E. Basar is with Communications Research and Innovation Laboratory (CoreLab), Department of Electrical and Electronics Engineering, Ko\c{c} University, Sariyer 34450, Istanbul, Turkey (e-mail: ebasar@ku.edu.tr)}
\thanks{M. Di Renzo is with Paris-Saclay University (L2S - CNRS, CentraleSup\'elec, University Paris Sud), Paris, France. (e-mail: marco.direnzo@l2s.centralesupelec.fr)}
\thanks{J. de Rosny is with Institut Langevin, Ecole Supérieure de Physique et de Chimie Industrielle, CNRS, 75005 Paris, France. (e-mail: julien.derosny@espci.fr)}
\thanks{M. Debbah is with CentraleSupelec, Universit\'e Paris-Saclay, 91192 Gif-sur-Yvette, France and Mathematical \& Algorithmic Sciences Lab, Huawei Technologies France SASU, 92100 Boulogne-Billancourt, France (e-mail: merouane.debbah@huawei.com)}
\thanks{M.-S. Alouini is with Computer, Electrical, and Mathematical Sciences and Engineering Division, King Abdullah University of Science and Technology, Thuwal 23955, Saudi Arabia (e-mail: slim.alouini@kaust.edu.sa)}
\thanks{R. Zhang is with Department of Electrical and Computer Engineering, National University of Singapore, Singapore 117583 (email: elezhang@nus.edu.sg)}
\thanks{The work of E. Basar is supported in part by the Scientific and Technological Research Council of Turkey (TUBITAK) under Grant 117E869, the Turkish Academy of Sciences (TUBA) GEBIP Programme, and the Science Academy BAGEP Programme.}  
\thanks{This paper has been presented in part at the 2019 European
Conference on Networks and Communications (EuCNC), Valencia,
Spain, June 2019 \cite{Basar_2019_LIS}.} 
}

\markboth{Basar \textit{et al.}: Wireless Communications Through Reconfigurable Intelligent Surfaces}
{Basar \textit{et al.}: Wireless Communications Through Reconfigurable Intelligent Surfaces}

\maketitle

\begin{abstract}
The future of mobile communications looks exciting with the potential new use cases and challenging requirements of future 6th generation (6G) and beyond wireless networks. Since the beginning of the modern era of wireless communications, the propagation medium has been perceived as a randomly behaving entity between the transmitter and the receiver, which degrades the quality of the received signal due to the uncontrollable interactions of the transmitted radio waves with the surrounding objects. The recent advent of reconfigurable intelligent surfaces in wireless communications enables, on the other hand, network operators to control the scattering, reflection, and refraction characteristics of the radio waves, by overcoming the negative effects of natural wireless propagation. Recent results have revealed that reconfigurable intelligent surfaces can effectively control the wavefront, e.g., the phase, amplitude, frequency, and even polarization, of the impinging signals without the need of complex decoding, encoding, and radio frequency processing operations. Motivated by the potential of this emerging  technology, the present article is aimed to provide the readers with a detailed overview and historical perspective on state-of-the-art solutions, and to elaborate on the fundamental differences with other technologies, the most important open research issues to tackle, and the reasons why the use of reconfigurable intelligent surfaces necessitates to rethink the communication-theoretic models currently employed in wireless networks. This article also explores theoretical performance limits of reconfigurable intelligent surface-assisted communication systems using mathematical techniques and elaborates on the potential use cases of intelligent surfaces in 6G and beyond wireless networks.
\end{abstract}


\IEEEpeerreviewmaketitle

\section{Introduction}
According to the February 2019 report of Cisco \cite{Cisco_2019}, by the year of 2022, the number of networked devices and connections will reach up to $ 28.5 $ billions, and $ 12.3 $ billions of them will consist of mobile-ready devices and connections. Furthermore, the overall mobile data traffic is expected to grow to $77$ exabytes per month by 2022 with a seven-fold increase over 2017. Meanwhile, after years of research and development, the first commercial 5th generation (5G) mobile communication standard (3GPP Release 15) has been completed in June 2018. As of mid-2019, 5G wireless networks have been already deployed in certain countries, while the first 5G compatible mobile devices are being introduced to the market. The advent of 5G has led to a new vision of mobile communications, which encompasses three use cases with different requirements: enhanced mobile broadband, ultra-reliable and low-latency communications, and massive machine type communications. However, one thing has become certain during the standardization process of 5G wireless networks: there is no single enabling technology that can support all 5G application requirements. From this perspective, researchers have already started research on beyond 5G, or even 6th generation (6G), technologies by escaping from the comfort-zone of 5G-oriented solutions. Even though future 6G technologies seem to be an extension of their 5G counterparts at present \cite{6G}, as 5G technologies were viewed 10 years ago, new user requirements, new applications and use cases, and new networking trends will bring more challenging communication engineering problems, which necessitate radically new communication paradigms, especially at the physical layer.

During the past few years, there has been a growing interest in novel communication paradigms in which the implicit randomness of the propagation environment is exploited to either simplify the transceiver architecture and/or to increase the quality of service (QoS). A notable example is spatial modulation (SM) \cite{SM_jour,SM_magazine,SM_magazine_2,Design_SM}, which is by far the most popular member of the index modulation (IM) family \cite{IM_5G,Basar_2017,IM_Book}. SM maps information bits onto transmit antenna indices by exploiting different fading realizations of multiple-input multiple-output (MIMO) antennas. Taking SM one step further, spatial scattering modulation \cite{SSM} and beam IM \cite{BIM} exploit the indices of the scatterers available in the environment to convey information. Finally, media-based modulation (MBM) utilizes reconfigurable antennas \cite{Khandani_conf1,MBM_TVT,Basar_2019} by encoding the information bits onto multiple distinguishable radiation patterns \cite{MDR-1,MDR-2,MDR-3}. In the aforementioned schemes, different signatures of the received signals, which originate from the interaction of the transmitted signals with the environment, are used to transmit information bits at a low implementation complexity. 

In the recent period, a brand-new technology was brought to the attention of the wireless research community: \textit{reconfigurable intelligent surfaces (RISs)}. The RISs are man-made surfaces of electromagnetic (EM) material that are electronically controlled with integrated electronics and have unique wireless communication capabilities. Current implementations include conventional reflect-arrays, liquid crystal surfaces, and software-defined meta-surfaces \cite{Akyildiz_2018_2,Lavigne_2018,Liu_2018}. In contrast to any other technology currently being used in wireless networks and  current design principles of wireless communications, the distinctive characteristic of RISs lies in making the environment controllable by the telecommunication operators, and by giving them the possibility of shaping and fully controlling the EM response of the environmental objects that are distributed throughout the network \cite{Akyildiz_2018}. The RISs share similarities but have major differences compared with SM-based systems: the RISs are, in fact, aimed to intentionally and deterministically control the propagation environment in order to boost the signal quality at the receiver \cite{Akyildiz_2018,Di_Renzo_2019,Basar_2019_LIS}. 

The RISs have given rise to the emerging concept of ``smart radio environments'' \cite{Di_Renzo_2019}. In contrast to current wireless networks where the environment is out of control of the telecommunication operators, a smart radio environment is a wireless network where the environment is turned into a smart reconfigurable space that plays an active role in transferring and processing information. Smart radio environments largely extend the notion of software networks. Future wireless networks, in particular, are rapidly evolving towards a software-based and reconfigurable platform, where every part of the network will be capable of adapting itself to the changes in the environment \cite{MDR-7}. In this optimization process, however, the environment itself remains an uncontrollable factor, i.e., it is unaware of the communication process undergoing within it.  Apart from being uncontrollable, the environment has usually a negative effect on the communication efficiency and the QoS. The signal attenuation limits the radio connectivity, multipath propagation results in fading phenomena, and reflections and refractions from large objects are the main sources of uncontrollable interference. In smart radio environments, on the other hand, the wireless environment itself is turned into a software-reconfigurable entity \cite{MDR-8}, whose operation is optimized to enable uninterrupted connectivity, high QoS guarantee, and where the information is transmitted without necessarily generating new signals but recycling the existing ones whenever possible \cite{Di_Renzo_2019}. 

RIS-empowered smart radio environments are a brand-new technology that has the potential of fundamentally changing how wireless networks are designed and optimized today. Controlling the propagation of radio waves opens the possibility of overcoming the negative effects of natural EM propagation, which is highly probabilistic in nature, by shaping how the radio waves interact with the surrounding objects that are coated with reconfigurable thin layers of EM material. In simple terms, the RISs allow network planners to counteract the destructive effect of multipath fading by coherently combining the radio waves reflected, refracted, and scattered from large surfaces \cite{Basar_2019_LIS}. The core technology behind this promising concept is the meta-surfaces, which is the 2D equivalent of meta-materials \cite{Akyildiz_2018,Di_Renzo_2019}.

It is worth noting that the RISs are different compared with other, and at the first sight, related technologies currently employed in wireless networks, such as relaying, MIMO beamforming, and backscatter communications. Details will be provided in the sequel, but it suffices to say that the RISs have the following distinguishable features:
\begin{itemize}
	
	\item They are nearly passive, and, ideally, they do not need any dedicated energy source.
	
	\item They are viewed as a contiguous surface, and, ideally, any point can shape the wave impinging upon it (soft programming).
	
	\item They are not affected by receiver noise, since, ideally, they do not need analog-to-digital/digital-to-analog converters (ADCs and DACs), and power amplifiers. As a result, they do not amplify nor introduce noise when reflecting the signals and provide an inherently full-duplex transmission.
	
	\item They have full-band response, since, ideally, they can work at any operating frequency.
	
	\item They can be easily deployed, e.g., on the facades of buildings, ceilings of factories and indoor spaces, human clothing, etc.

\end{itemize}

These distinctive characteristics make RIS-assisted communication a unique technology, but introduce important design challenges, which will be discussed and elaborated in the sequel.

Although traditional meta-surfaces with fixed EM functionalities have been used in various applications, including radar and satellite communications, their application in mobile communications is relatively limited \cite{Wu_2019}. Passive surfaces that do not have the ability to alter (or reconfigure) their EM characteristics to control the propagation environment, in fact have a very limited impact in highly dynamic wireless communication environments. For application to wireless networks, it is fundamental that the meta-surfaces are reconfigurable in order to adapt themselves according to the changes of the wireless environment \cite{Akyildiz_2018_2}. The modeling, analysis, and design of RISs for application to wireless networks is a highly multidisciplinary research endeavor at the intersection of wireless communications, communication theory, computer science, physics, electromagnetism, and mathematics. Within this context, we may need to revisit meta-materials from the perspective of communication engineering by combining physical and digital domains.

The aim of the present article is to summarize the latest research activities on RIS-empowered wireless networks, to elaborate on the fundamental differences with other technologies, to discuss the most important open research issues to tackle, and to highlight why the use of RISs necessitates to rethink the communication-theoretic models currently employed in wireless networks. We will describe, in particular, the potential use of RISs either for sharping the radio waves or for realizing low-complexity MIMO transmitters. Simple analytical models to convey the distinguishable features of RISs will be used, with focus on link budget (path-loss) and error performance.

The rest of the article is summarized as follows. We revisit the popular two-ray model in wireless communications to illustrate the concept of controllable wireless propagation in Section II. In Section III, we shed light on the basic operation mechanisms of RISs. In Section IV, we provide a mathematical framework for the calculation of the error performance of RIS-assisted systems. RISs employed as low-complexity transmitters are introduced in Section V. A detailed historical perspective and an overview of state-of-the-art solutions on RISs are presented in Section VI. Potential use cases and open research issues are discussed in Sections VII and VIII, respectively. Finally, Section IX concludes the paper.

\section{Controllable Wireless Propagation Through Reconfigurable Intelligent Surfaces -- An Illustrative Example}

In a typical wireless communication environment, a transmitted radio signal encounters multiple objects on its way, which produce reflected, diffracted, and scattered replicas of the transmit signal. These copies are called multipath components, and arrive at the receiver with  different (most probably random and unpredictable) magnitudes, phases, and delays that produce significant distortions on the received signal because of their constructive and destructive summation. This effect is known as fading in wireless communications and is a major limiting factor in modern and future wireless communication systems. The main motivation of using RISs is to realize a controllable radio environment, in which the highly probabilistic wireless channel is turned into a deterministic space by carefully re-engineering the propagation of the EM waves in a software-controlled fashion. In this section, we illustrate the basic working mechanism of RISs by focusing on their use to modify the signals reflected by large planar surfaces. We consider a simple example that is based on revisiting the well-known two-ray channel model in a free-space environment, but in the presence of an RIS deployed on the ground plane.

\begin{figure}[!t]
	\begin{center}
		\includegraphics[width=1\columnwidth]{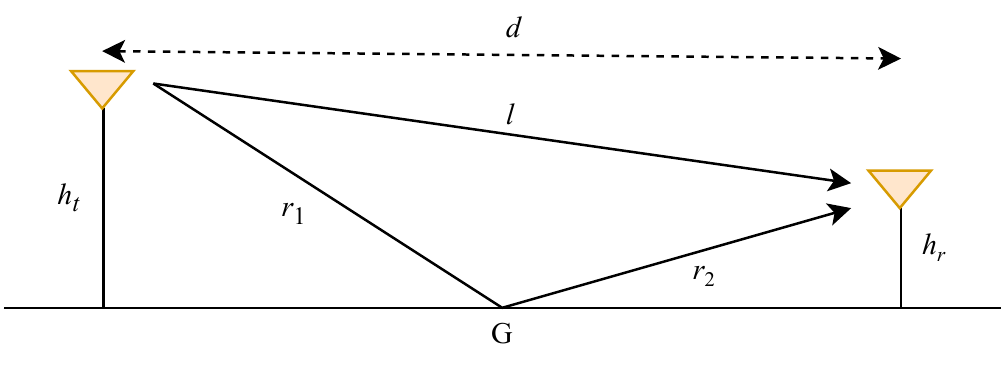}
		\caption{Two-ray propagation model with a LOS ray and a ground-reflected ray.}
		\label{fig:ground_reflection}
	\end{center}
\end{figure}

\subsection{The Conventional Two-Ray System Model}

In particular, we consider the two-ray channel model depicted in Fig. \ref{fig:ground_reflection}. In this model, the received signal consists of two components: the line-of-sight (LOS) ray and the ray reflected from the ground. Our system model and analysis are based on a geometrical optics (or ray optics) model for the propagation of radio waves \cite{MDR-9}. Geometrical optics, or ray optics, is a model of optics that describes the propagation of radio waves in terms of rays, and assumes that the geometric size of the objects is much larger than the wavelength of the radio wave. For ease of illustration, we assume that the ground plane is sufficiently large compared with the transmission wavelength and that it produces only specular reflections. Further information on the geometric size of the RISs to behave as reflectors is given in further text. Under these assumptions, the energy is regarded as being transported along certain curves, i.e., the radio waves (the rays) are assumed to propagate in straight-line paths if they travel in homogeneous media, and to bend and split in reflected and refracted signals at the interface between two dissimilar media. In more general terms, the propagation of radio waves modeled as rays adhere to the Fermat's principle, which states that the path taken by a ray between two points is the path that is traversed in the least time.

In Fig. \ref{fig:ground_reflection}, the distance between the transmit and receive antennas is denoted by $l$, and the distance between the point of reflection, G, on the ground and the transmit and receive antennas are denoted by $r_1$ and $r_2$, respectively. According to the geometrical optics and Fermat's principle, the point of reflection G corresponds to the trajectory that allows the transmitted signal to reach the receiver in the least time. This is the well-known Snell's law of reflection. Accordingly, G is the point in which the angle between the incident ray and the perpendicular line to the ground is the same as the angle between the reflected ray and the perpendicular line to the ground \cite{MDR-10}.

We denote the ground reflection coefficient by $R$, which typically depends on the properties of the material of the ground, the polarization of the radio wave, and the angle of incidence \cite{Goldsmith}. Without loss of generality, we assume unit gain transmit/receive antennas and a narrow-band transmission signal, i.e., $x(t)\approx x(t-\tau)$, where $x(t)$ is the complex baseband transmitted signal and $\tau$ is the relative time delay between the ray reflected from the ground and the LOS path, which is given by $\tau=(r_1+r_2-l)/c$ with $c$ being the speed of light. Then, the received (noise-free) baseband signal can be expressed as follows:
\begin{equation}
\label{1}
r(t)= \frac{\lambda}{4\pi}  \left(\frac{e^{-\frac{j2\pi l}{\lambda}}}{l} + \frac{R \times e^{-\frac{j2\pi (r_1+r_2)}{\lambda}}}{r_1+r_2} \right) x(t)
\end{equation}
where $\lambda$ is the wavelength. In simple terms, the received signal is the sum of the LOS and ground-reflected signals with phase delays $2\pi l/ \lambda$ and $2\pi (r_1+r_2)/ \lambda$, respectively, which are proportional to the propagation distances.  

Assuming that the transmit power of $x(t)$ is $P_t$, the received power $P_r$ can be expressed, from \eqref{1}, in terms of $P_t$ as follows:
\begin{equation}
\label{2}
P_r=P_t \left(\frac{\lambda}{4\pi} \right)^2 \left| \frac{1}{l} + \frac{R\times e^{-j\Delta\phi}  }{r_1+r_2} \right|^2
\end{equation}
where $\Delta\phi=\frac{2\pi (r_1+r_2-l)}{\lambda}$ is the phase difference between the two paths. 

Assuming that the distance $d$ is large enough, i.e., $d \gg h_t+h_r$, then we obtain $d \approx l \approx r_1+r_2$ and $R\approx -1$ for a specular reflection from the ground \cite[Eq. (2.15)]{Goldsmith}. Therefore, \eqref{2} simplifies as follows:
\begin{equation}
P_r\propto P_t \left(\frac{1}{d^2} \right)^2
\end{equation}
which decays with the fourth power of the distance $d$. 

If the ground reflection is not present, i.e., the second term in \eqref{1} is equal to zero, the LOS free-space propagation model yields a received signal power that decays with the second power of the distance:
\begin{equation}
P_r=P_t \left(\frac{\lambda }{4\pi d}\right)^2.
\end{equation}

Comparing (3) and (4), one can easily observe the destructive effect, on the power of the received signal, that the uncontrollable reflection from the ground generates because of the misaligned phases of the two paths shown in Fig. \ref{fig:ground_reflection}. In other words, just a single uncontrollable reflection from the ground may cause major degradations on the received signal strength, even in the very optimistic transmission scenario with no user mobility and no random effects induced by the environment.   

\subsection{The Two-Ray System Model with a Single Reconfigurable Meta-Surface}

Let us consider the same system model with the only exception that a reconfigurable meta-surface is laid on the ground to assist the communications between the transmitter and receiver. In particular, we assume the simple case study where the meta-surface acts as a reflecting surface, which is capable of modifying the direction of the reflected ray (i.e., the angle of reflection) according to the generalized Snell's law \cite{MDR-10}, as well as the phase of the reflected ray as described in \cite{MDR-11} and \cite{MDR-12}. Further details on the operation of reconfigurable meta-surfaces are provided in the next section. It suffices to say that the angle and phase of the reflected ray can be modified by engineering the phase gradient of the meta-surface \cite{MDR-10,MDR-11,MDR-12}. Similar to the reflection from the ground, the reflection coefficient usually depends on the characteristics of the incident wave, e.g., the polarization of the incident EM field, the material that the meta-surface is made of, and the angles of incidence and reflection \cite{MDR-11}.

In the considered example, we focus our attention on the possibility of optimizing the phase of the reflected ray and assume that no anomalous reflection is needed, i.e., the Snell's law applies. Also, we assume that the entire ground is coated with a reconfigurable meta-surface. Conceptually, the reconfigurable meta-surface can be viewed as an ideal phase shifter that is capable of adjusting the phase of the reflected wave so that the LOS and reflected rays sum up coherently, and the signal strength of their sum is maximized. If we assume that the reconfigurable meta-surface is capable of optimizing the phase of the reflected ray in an optimal fashion, i.e., by coherently aligning the phases of the direct and the reflected rays for any angles of incidence and reflection, we would obtain the following:
\begin{equation}
P_r=P_t \left(\frac{\lambda}{4\pi} \right)^2 \left| \frac{1}{l} + \frac{1  }{r_1+r_2}  \right|^2 \approx 4 P_t \left(\frac{\lambda }{4\pi d}\right)^2
\end{equation}
which corresponds to setting $R=e^{j\Delta\phi}$, and by considering $d \approx l \approx r_1+r_2$.

By comparing (3) with (5), we evince that the use of reconfigurable meta-surfaces has the potential of changing the scaling law that governs the received power as a function of the distance: the received power does not decay anymore with the forth power of the distance but only with the second power of the distance, which is the same as the LOS ray. In further text, we will show that this simple result is one of the main distinctive differences of RISs with respect to relaying and backscatter communications. 

\begin{figure}[!t]
	\begin{center}
		\includegraphics[width=1\columnwidth]{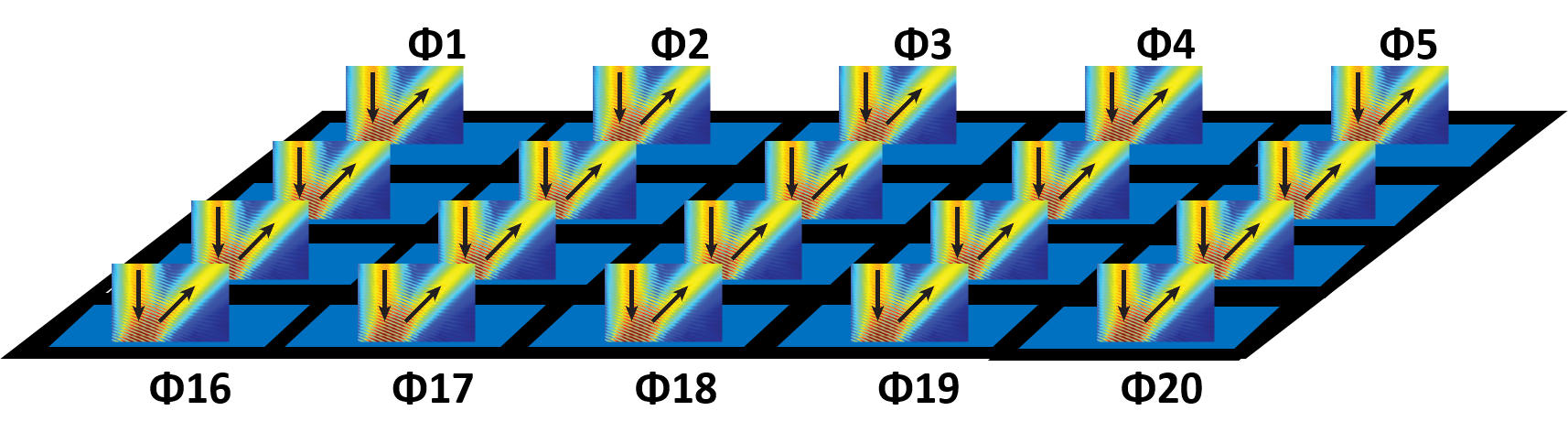}
		\caption{Conceptual illustration of an RIS made of 20 reconfigurable meta-surfaces whose phases ($\Phi_i$) can be tuned independently and whose reflected rays are steered toward the same reflection angle. For ease of representation, multiple incident rays are depicted, but only one incident ray is present in practice.}
		\label{fig:Fig_MDR}
	\end{center}
\end{figure}

\subsection{The Two-Ray System Model with An RIS Made of Many Reconfigurable Meta-Surfaces}

Let us now go one step further, and assume that the ground floor is not coated with a single reconfigurable meta-surface but with an RIS that is made of $N$ reconfigurable meta-surfaces each of which capable of tuning the angle of reflection according to the Snell's law and the phase of the reflected ray independently of the other meta-surfaces. A conceptual diagram of the considered system model is sketched in Fig. \ref{fig:Fig_MDR}. In the sequel, we will keep using the following terminology: i) a reconfigurable (reflecting) meta-surface is a surface that allows the angle and the phase of the reflected ray to be arbitrarily controlled, and ii) an RIS is the collection of several reconfigurable (reflecting) meta-surfaces that are capable of steering the reflected ray towards the same angle and of adjusting the phase of the reflected ray independently of the other reconfigurable meta-surfaces. Under these assumptions, the received signal power can be written as follows:
\begin{equation}
\label{eq:6}
P_r=P_t \left(\frac{\lambda}{4\pi} \right)^2 \left| \frac{1}{l} + \sum_{i=1}^{N} \frac{R_i\times e^{-j\Delta\phi_i}  }{r_{1,i}+r_{2,i}} \right|^2
\end{equation}
where the symbols have the same meaning as in \eqref{2}, and the index $i$ is referred to the $i$th reconfigurable meta-surface of the RIS. 

From \eqref{eq:6}, we evince that the power of the received signal may fluctuate significantly if the reflection coefficients of the $N$ reconfigurable meta-surfaces are not optimized. Let us assume that each $R_i$ is optimized so that the phase of the received signal from the $N$ reflecting meta-surfaces is aligned with the phase of the LOS path, i.e., $R_i=e^{j\Delta\phi_i}$ and $r_{1,i}+r_{2,i}\approx l \approx d$ for all $i$. Then, the received power can be formulated as follows:
\begin{equation}
\label{eq:7}
P_r \approx (N+1)^2  P_t \left(\frac{\lambda }{4\pi d}\right)^2.
\end{equation}

By direct inspection of \eqref{eq:7}, two major conclusions can be drawn: 1) the received power is proportional to $N^2$, which is the number of phases of the RIS that can be controlled independently, and 2) the received power decays with the inverse of the square of the distance between the transmitter and the receiver. In other words, as a function of the distance, the received power decays as for the LOS path, while a power gain that is proportional to the square of the number of controllable phases is obtained. This shows the potential of using the RISs in wireless networks. It is worth mentioning that this illustrative example is relatively simple and relies on a number of assumptions, e.g., the possibility of optimizing the reflection phases without any discretization error and for any angles of incidence and reflection, the absence of reflection losses, and the perfect knowledge of the phases at the RIS. In addition, the optimization of the phases is usually not a straightforward task for more practical system models, e.g., \cite{Huang_2019}.

\subsection{On the Geometric Size of an RIS to be a Specular Reflector}

From \eqref{eq:7}, it is worth investigating the size that an RIS is expected to have as a function of the number of reflecting meta-surfaces that it is made of, i.e., $N$. As a reference for this study, we can consider the samples of meta-surfaces available in \cite{MDR-13,MDR-14,MDR-15,MDR-16}. Based on, e.g., \cite{MDR-13}, we can assume that a meta-surface that is capable of shaping the angle and the phase of the reflected signal has a size of the order of $10\lambda  \times 10\lambda$. This size allows, in general, a meta-surface to be viewed as a specular reflector according to geometrical optics. If we assume that an RIS is made of $N=100$ reconfigurable meta-surfaces whose reflection phase can be controlled independently of the others, then the size of the RIS would be of the order of $100\lambda  \times 100\lambda$. If the operating frequency is of the order of $ 30 $ GHz, e.g., for application in the millimeter-wave frequency band where the RISs may have promising applications for enhancing the coverage in cellular networks as discussed in the sequel, then we have $\lambda  \approx 1$ cm, which results in an RIS of 1 m$^2$. A structure of this type can be readily deployed either in indoors or outdoors, and yields, according to \eqref{eq:7}, the same power decay as a function of the distance as the LOS path, but a $\sim N^2 = 100^2 = 10^4 = 40$ dB increase of the received power. Even though, as just mentioned, these results are obtained under a number of ``comfortable'' assumptions, the potential gain is sufficiently large to motivate further research on the potential and limitations of using the RISs in wireless networks.

\subsection{Intelligent Reflection vs. Relaying and Backscattering: Reflectors vs. Diffusers}

We close this section, by elaborating on the main difference that renders the concept of RISs a unique and peculiar technology when compared to, at the first sight, similar technologies. The two technologies that are often deemed to be equivalent to the RISs are relay-aided transmission \cite{MDR-18} and backscatter communications \cite{MDR-19}. As far as the relays are concerned, in particular, the most similar approach is the amplify-and-forward (AF) or transparent relaying scheme \cite[Sec. 2.4]{MDR-18}. From \cite[Eq. (2.118)]{MDR-18} and \cite[Eqs. (2)-(4)]{MDR-19}, we evince that the received power decays with the fourth power of the distance for transparent relaying and backscatter communications. By ignoring the LOS path, in particular, the received power can be formulated as follows:
\begin{equation}
P_r \propto P_t \left( \frac{1}{r_1}\right) ^2  \left( \frac{1}{r_2}\right) ^2
\end{equation}
where $r_1$ and $r_2$ stand for the distances between the relay and the terminals. By assuming $r_1 \approx r_2 \approx d/2$, i.e., the relay is placed mid-way between the transmitter and the receiver, we obtain $P_r \propto P_t / d^4$ as reported in \cite[Eq. (2.118)]{MDR-18} and \cite[Eqs. (2)-(4)]{MDR-19}. This is the well-known ``product channel'' or radar-like equation that is usually used for analyzing relaying and backscatter communications. Based on the above product channel model, a comprehensive overview of RIS-assisted wireless networks can be found in \cite{Wu_2019}.

By direct inspection of \eqref{eq:7}, we note that the scaling law of the received power as a function of the distance of the RISs is in sharp contrast with the scaling law that dictates the path-loss of relay-aided and backscatter communications. This is a simple but net evidence of the difference between the RISs and the two most similar approaches currently available in the literature. The motivation of the different scaling law lies in the geometric size of the RISs, the geometric size of the antennas that constitute the relays (even if arrays of antennas are employed), and the geometric size of the tags used for backscattering. The geometric size of the RISs is large enough, i.e., much larger than the wavelength, to be modeled as specular reflectors. The geometric size of the antennas of relays and backscattering tags is, on the other hand, smaller than or comparable with the wavelength, which render AF relaying and backscatter tags diffusers rather than specular reflectors. The different path-loss function of the RISs is due, in addition, to their passive nature, i.e., they are supposed not to store and process the impinging signals. The geometric size of the RISs, if appropriately optimized, may yield remarkable link budget gains in comparison with relaying and backscatter communications. It is worth mentioning, in addition, that a perfect RIS, i.e., with infinite size and no reflection losses, in the presence of transmitters and receivers with omni-directional antennas is capable of focusing towards the receiver half of the transmit power regardless of the distance.

Finally, we emphasize that the path-loss model in (1) is widely used in ray tracing in order to model specular reflections, e.g., \cite{Landron_1996}. The difference between the specular reflections in (1) and the diffusely scattered field in (8) is briefly discussed in \cite{Schaubach_1992}. The path-loss model in (1) was recently employed in \cite{Mehrotra_2019} to analyze the performance of RISs for application to millimeter-wave communications.

\section{Reconfigurable Intelligent Surfaces: How Do They Work?}

The RISs are reconfigurable sheets of EM material that intentionally control the propagation in the environment in order to enhance the signal quality at the receiver. The RISs are made of a large number of low-cost and passive elements that are capable of modifying the radio waves impinging upon them in ways that naturally occurring materials are not capable of. A simple example where the RIS is made of meta-surfaces that act as programmable reflectors is depicted in Fig. \ref{fig:Fig_MDR}. Unlike other similar technologies, e.g., relays and MIMO beamforming, the RISs do not require any power source and complex processing, encoding, and decoding algorithms. It is worth mentioning that the RISs are often referred to as software-defined surfaces (SDSs) in analogy with the concept of software-defined radio (SDR), i.e., ``a radio in which some or all of physical layer functions are software defined''. According to this terminology, an RIS can be viewed as an SDS whose response of the radio waves is programmed in software. In this section, we describe the operating principle of the RISs in simple but general terms.

\begin{figure}[!t]
	\begin{center}
		\includegraphics[width=0.83\columnwidth]{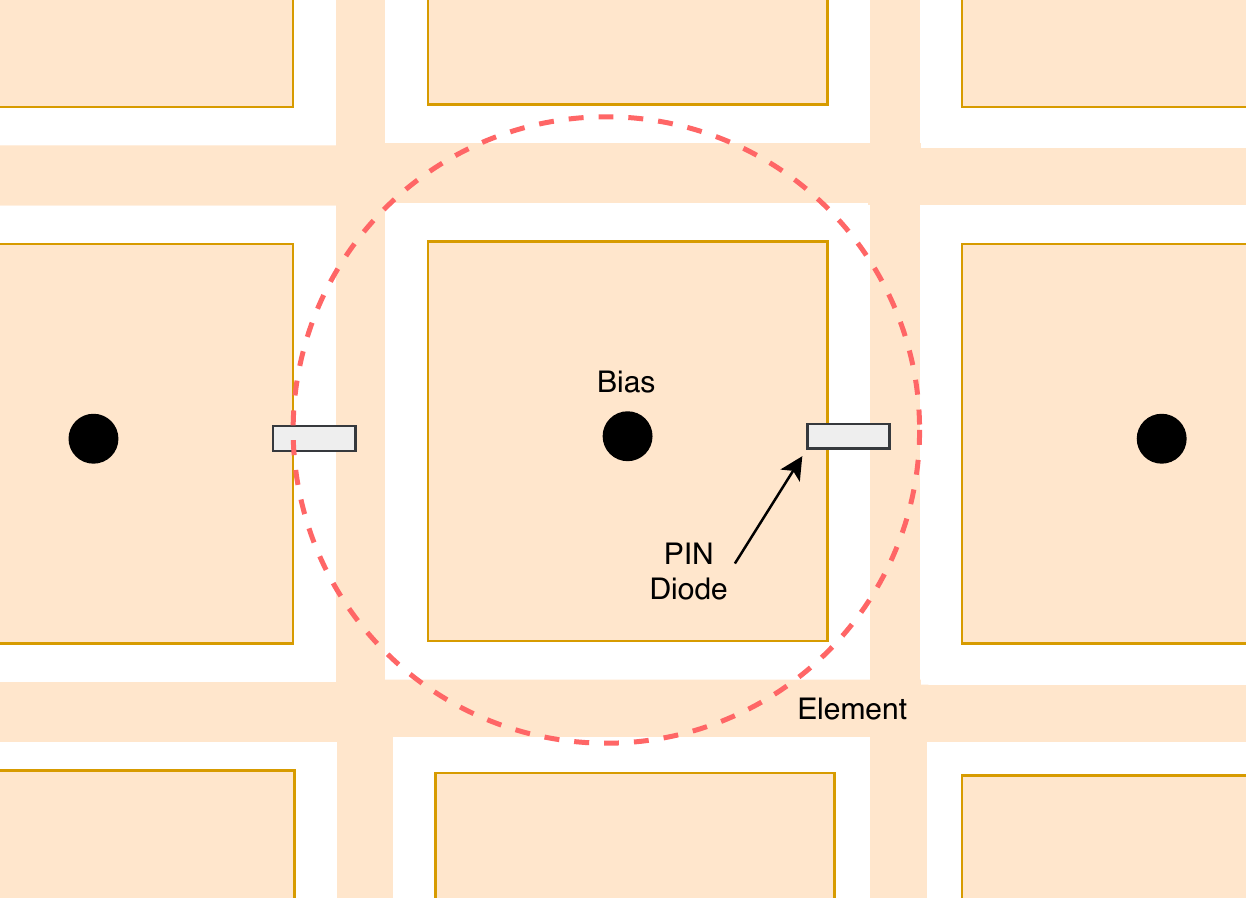}
		\caption{The structure of an RIS (intelligent wall) that is made of an active frequency-selective surface.}
		\label{fig:intelligent_wall}
	\end{center}
\end{figure}

In \cite{Subrt_2012}, the authors introduced intelligent walls that are equipped with frequency-selective surfaces. These surfaces have a planar structure and PIN diodes are embedded on the metal connection parts of each surface element. These PIN diodes are switched on and off by an external bias and provide two different states for the intelligent wall. In the first state (when the PIN diodes are off), an almost transparent surface, which allows the incoming energy to pass through, is obtained. When the PIN diodes are switched to the second state, on the other hand, the majority of the incident energy is reflected. In other words, two important EM functionalities (the waves either pass through or are reflected from the surface) are realized by an intelligent wall. In Fig. \ref{fig:intelligent_wall}, the structure of this intelligent wall is illustrated.

In \cite{MDR-XX}, the authors fabricated a $ 0.4 $ m$^2$ spatial microwave modulator that consists of $ 102 $ controllable EM reflectors, and operates at a working frequency of $ 2.47 $ GHz. These $ 102 $ reflectors are controlled by using two Arduino $ 54 $-channel digital controllers. The authors demonstrated that spatial microwave modulators can efficiently shape, in a passive way, complex existing microwave fields in reverberating environments with a non-coherent energy feedback. In particular, the authors showed that binary-only phase state tunable meta-surfaces allow one to obtain a good control of the waves, owing to the random nature of the EM fields in complex media. Subsequent research works from the same group of researchers can be found in \cite{MDR_A,MDR_B}.

\begin{figure}[!t]
	\begin{center}
		\includegraphics[width=1\columnwidth]{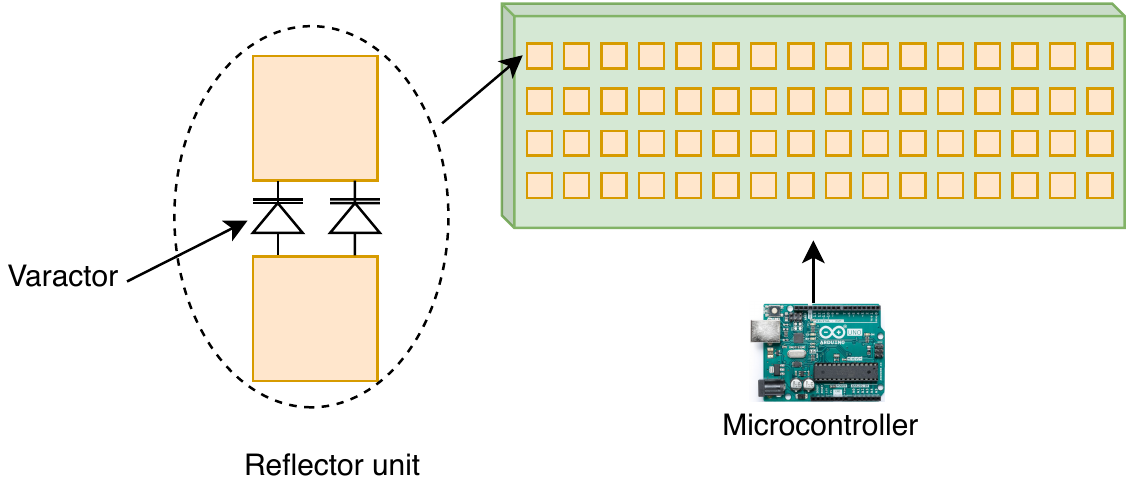}
		\vspace*{-0.35cm}\caption{Reconfigurable reflect-arrays with tunable resonators.}\vspace*{-0.4cm}
		\label{reflectarray}
	\end{center}
\end{figure}

Another approach for obtaining reconfigurable and smart reflect-arrays is to use varactor-tuned resonators \cite{Hum_2014}, as shown in Fig. \ref{reflectarray}. The idea of this implementation is to change the resonant frequency of the available patches by electronic tuning instead of changing the resonator dimensions as done in non-reconfigurable reflect-arrays. In this setup, a tunable capacitor (varactor) is used in each reflector unit and a tunable phase shift is obtained by adjusting the bias voltage applied to the varactor in order to change its capacity. Using this approach, a smart reflector with $ 48 $ patch elements is constructed in \cite{Tan_2016}. In particular, the EM response of the patch elements can be altered by using micro-controllers, which generate input signals to tune the varactors and to change the phase of the reflected signal. A more advanced reflect-array with  $224$ reconfigurable patches is designed in \cite{Tan_2018} for application to $60$ GHz WiFi signals, which is made of electronically-controlled relay switches.  In this implementation, each reflector can be turned on and off according to the status of its switch. A beam searching-based reflect-array control algorithm is introduced as well, where the access point (AP) and the reflect-array perform beam searching to ensure a maximized signal quality at the intended user. Due to hardware limitations, however, a binary phase control (two possible phases) is considered, which causes a degradation of the received signal-to-noise ratio (SNR).

The concept of HyperSurfaces is based on coating objects, such as walls or furniture, with thin sheets of EM material  that enable one to control the EM behavior of a wireless environment through software \cite{Akyildiz_2018}. The HyperSurfaces belong to the family of software-controlled meta-surfaces. Depending on the states of electronic switches that are embedded throughout the meta-surface, the distribution of the current can be controlled, which enables the meta-surface to adapt its response depending on the impinging radio wave and on the desired response. From this perspective, one can easily notice the conceptual similarity between this implementation of a meta-surface and reconfigurable antennas, in which the resulting radiation pattern is altered by changing the current distribution. In \cite{Akyildiz_2018}, the constituent meta-surfaces depicted in Fig. \ref{fig:Fig_MDR} are referred to as ``tiles'', which are rectangular structures that can realize functions such as wave steering, wave polarization, and wave absorption, in a software-defined fashion. In simple terms, a HyperSurface tile supports various software-defined EM functions, which can be programmed in software by setting the direction of the incident wave, the direction of the intended reflection, and the frequency band of interest, etc.

Liquid-crystal reconfigurable meta-surface-based reflectors are proposed in \cite{Foo_2017} by exploiting electronically tunable liquid crystals to enable the real-time reconfigurability of the meta-surfaces for beam steering. By varying DC voltages on microstrip patches of liquid crystal loaded unit cells, the effective dielectric constant of each unit can be adjusted. Consequently, the phase shifts at various locations of a meta-surface can be controlled in real-time and the reflected wave can be manipulated.

\section{Controlling the Multipath Through Reconfigurable Intelligent Surfaces}
\begin{figure}[!t]
	\begin{center}
		\includegraphics[width=1\columnwidth]{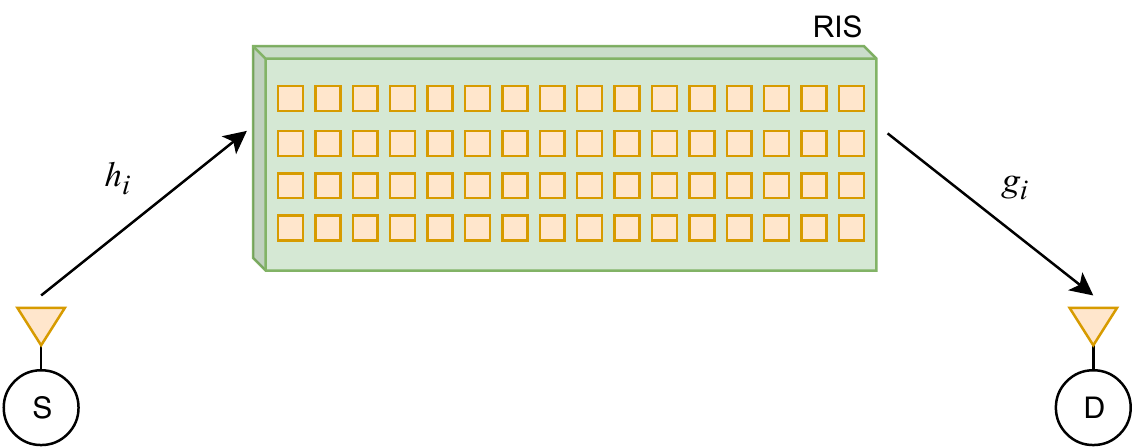}
		\caption{Reflection from an RIS in a dual-hop communication scenario without a line-of-sight path between S and D.}
		\label{fig:LIS_DH}
	\end{center}
\end{figure}

In this section, we present the system model of a generic RIS-based single-input single-output (SISO) scheme and introduce a unified framework for the calculation of the symbol error probability (SEP) through the derivation of the received SNR distribution. The block diagram of the considered RIS-based transmission scheme is shown in Fig. \ref{fig:LIS_DH}, where $h_i$ and $g_i$ are the fading channels between the single-antenna source (S) and the RIS, and between the RIS and the single-antenna destination (D) for the $i$th reflecting meta-surface $(i=1,2,\ldots,N)$, and $N$ is the number of reflecting meta-surfaces of the RIS. Under the assumption of Rayleigh fading channels, we have $h_i,g_i \sim \mathcal{CN}(0,1)$, where $\mathcal{CN}(0,\sigma^2)$ stands for complex Gaussian distribution with zero mean and $\sigma^2$ variance. For clarity, we emphasize that, as usual practice, the path-loss is not considered in the fading coefficients $ h_i $ and $ g_i $, since it is implicitly taken into account in the (receiver) SNR that is defined in further text. Therefore the structure of the RIS is similar to that depicted in Fig. \ref{fig:Fig_MDR}, and we assume that it provides adjustable phase shifts that are controlled by and programmed through a communication-oriented software. In our analysis, we assume perfect knowledge of the channel phases of $h_i$ and $g_i$ for  $i=1,2,\ldots,N$ at the RIS, which corresponds to the best scenario in terms of system operation and yields a performance benchmark for practical applications.

Let us assume a slowly varying and flat fading channel model. The received signal reflected by an RIS that is made of $N$ reflecting meta-surfaces can be expressed as follows:
\begin{equation}\label{eq:1}
r=\left[ \sum_{i=1}^{N} h_i e^{j \phi_i} g_i \right] x + n 
\end{equation}
where $\phi_i$ is the adjustable phase induced by the $i$th reflecting meta-surface of the RIS, $x$ stands for the data symbol selected from an $M$-ary phase shift keying/quadrature amplitude modulation (PSK/QAM) constellation and $n \sim \mathcal{CN}(0,N_0)$ is the additive white Gaussian noise (AWGN) sample. As far as the channels are concerned, we have $h_i = \alpha_i e^{-j \theta_i}$ and $g_i = \beta_i e^{-j \psi_i}$. In matrix form, \eqref{eq:1} can be also re-written as follows:
\begin{equation}\label{eq:matrix}
r= \mathbf{g}^{\mathrm{T}} \Phi \mathbf{h} x +n
\end{equation} 
where $\mathbf{h}=\begin{bmatrix}
h_1 & h_2 & \ldots & h_N
\end{bmatrix}^{\mathrm{T}}$ and $\mathbf{g}=\begin{bmatrix}
g_1 & g_2 & \ldots & g_N
\end{bmatrix}^{\mathrm{T}}$ represent the vectors of channel coefficients between the two terminals and the RIS, and $\Phi=\mathrm{diag} ( \left[ e^{j\phi_1} \,\, e^{j\phi_2}\,\,\ldots\,\,e^{j\phi_N} \right] )$ is a diagonal matrix that contains the phase shifts applied by the reflecting meta-surfaces of the RIS. By direct inspection of \eqref{eq:matrix}, the obtained model resembles that of a precoding/beamforming system in conventional MIMO systems. However, the precoding/beamforming operation is realized over the transmission medium (the environment) rather than at the transmitter or at the receiver. As far as the fast fading is concerned, we note, in particular, that the channel model for the RISs and the relays are similar, while their path-loss models are different if the RISs are sufficiently large to behave as reflectors.

From \eqref{eq:1}, the instantaneous SNR at D can be formulated as follows:
\begin{equation}
\gamma =  \frac{\left|  \sum_{i=1}^{N} \alpha_i \beta_i e^{j (\phi_i-\theta_i - \psi_i   ) } \right|^2 E_s }{N_0} 
\end{equation}
where $E_s$ is the average transmitted energy per symbol. It is not difficult to infer that $\gamma$ is maximized by eliminating the channel phases (similar to co-phasing in classical maximum ratio combining schemes), i.e., the optimal choice of $\phi_i$ that maximizes the instantaneous SNR is $\phi_i = \theta_i + \psi_i$ for $i=1,\ldots,N$. This solution, notably, requires that the channel phases are known to the RIS. How to perform channel estimation in RIS-empowered wireless networks along with the challenges that this entails if the RISs are assumed to be passive, as opposed to, e.g., the relays, is discussed in the sequel. In detail, the optimal choice of the phases, $\phi_i = \theta_i + \psi_i$ for $i=1,\ldots,N$, is obtained from the identity:
\begin{equation}
\left| \sum_{i=1}^{N} z_i e^{j \xi_i}\right| ^2 = \sum_{i=1}^{N} z_i^2 + 2 \sum_{i=1}^{N} \sum _{k=i+1}^N  z_i z_k \cos(\xi_i - \xi_k) 
\end{equation}
which is maximized if $\xi_i =\xi$ for all $i$. 

Therefore, the maximized SNR can be formulated as follows:
\begin{equation}
\gamma =\frac{\left( \sum_{i=1}^{N} \alpha_i \beta_i   \right)^2 E_s }{N_0}= \frac{ A^2 E_s }{N_0}.
\end{equation}

Since $\alpha_i$ and $\beta_i$ are independently Rayleigh distributed random variables (RVs), the mean value and the variance of their product are $\mathrm{E}[\alpha_i \beta_i]= \frac{\pi}{4}$ and $ \mathrm{VAR}[\alpha_i \beta_i] = 1-\frac{\pi^2}{16}$, respectively. For a sufficiently large number of reflecting meta-surfaces, i.e., $N \gg 1$, according to the central limit theorem (CLT), $A$ converges to a Gaussian distributed random variable with parameters $\mathrm{E}[A]=\frac{N \pi}{4} $ and $\mathrm{VAR}[A]=N\left(1-\frac{\pi^2}{16} \right) $. Therefore, $\gamma$ is a non-central chi-square random variable with one degree of freedom and has the following moment generating function (MGF) \cite{Proakis}:
\begin{equation}\label{MGF_1}
M_{\gamma}(s)= \left( \dfrac{1}{1-\frac{sN(16-\pi^2)E_s}{8N_0}}\right) ^{\frac{1}{2}}  \! \!\exp\left( \dfrac{ \frac{sN^2 \pi^2 E_s}{16 N_0}}{1-\frac{sN(16-\pi^2)E_s}{8N_0}} \right). 
\end{equation}

Furthermore, the average received SNR can be written as follows:
\begin{equation}\label{key}
\mathrm{E}\left[ \gamma \right]=\frac{( N^2 \pi^2+ N(16-\pi^2)) E_s}{16N_0}
\end{equation}
which is proportional to $N^2$, i.e., $\mathrm{E}\left[ \gamma \right]\propto N^2 \frac{E_s}{N_0}$, if $N \gg 1$. This result is in agreement with the received signal power in \eqref{eq:7}.

From \eqref{MGF_1}, we can compute the average SEP of $M$-ary phase shift keying (PSK) signaling as follows \cite{Simon}:
\begin{equation}\label{5}
P_e=\frac{1}{\pi} \int_{0}^{(M-1)\pi/M} M_{\gamma} \left( \frac{-\sin^2 (\pi/M)}{\sin^2 \! \eta}\right) d\eta
\end{equation}
which for binary PSK (BPSK) simplifies to
\begin{align}\label{SEP_1}
	P_e= \frac{1}{\pi}  \! \int_{0}^{\pi/2} \! \!& \left( \dfrac{1}{1+\frac{N(16-\pi^2)E_s}{8 \sin^2 \! \eta N_0}}\right) ^{\!\frac{1}{2}}   \nonumber \\
	&\times \exp\left( \dfrac{ -\frac{N^2 \pi^2 E_s}{16  \sin^2 \! \eta N_0 }}{1+\frac{N(16-\pi^2)E_s}{8 \sin^2 \! \eta N_0}} \right) \! \! d\eta. 
\end{align}

In order to gain further insights, \eqref{SEP_1} can be upper bounded by setting $\eta=\pi/2$, which yields:
\begin{equation}\label{SEP_1_UB}
P_e \le \frac{1}{2}   \left( \dfrac{1}{1+\frac{N(16-\pi^2)E_s}{8N_0 }}\right) ^{\!\frac{1}{2}}  \! \!\! \exp\left( \dfrac{ -\frac{N^2 \pi^2 E_s}{16 N_0 }}{1+\frac{N(16-\pi^2)E_s}{8N_0}} \right). 
\end{equation}

\begin{figure}[!t]
	\begin{center}
		\includegraphics[width=1\columnwidth]{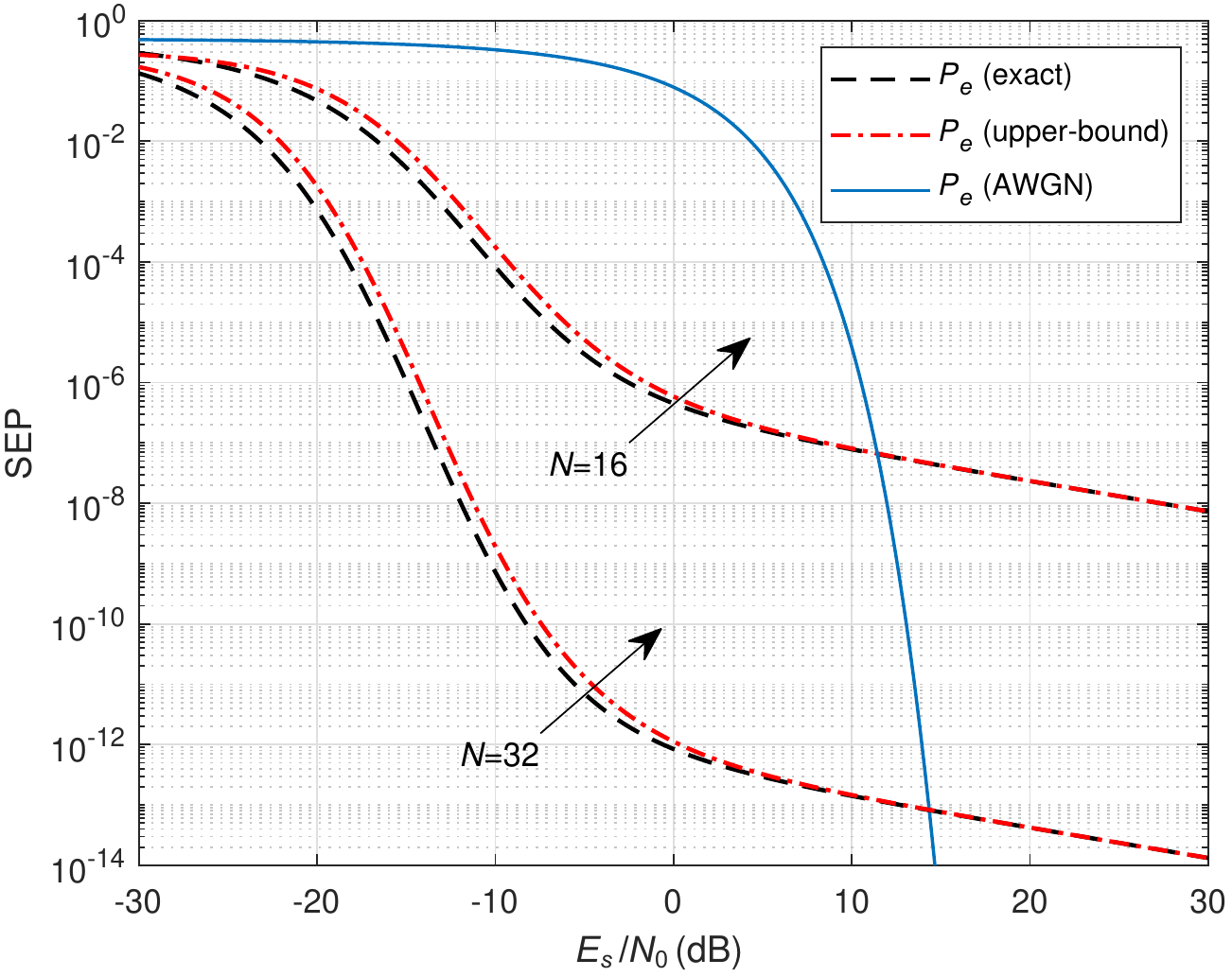}
		\caption{Theoretical average SEP of RIS-based scheme for $N=16$ and $N=32$ with BPSK.}
		\label{fig:N_16_32_BER}
	\end{center}
\end{figure}

In Fig. \ref{fig:N_16_32_BER}, we plot the average SEP of the considered RIS-based scheme by using \eqref{SEP_1} and \eqref{SEP_1_UB} for $N=16$ and $N=32$. From Fig. \ref{fig:N_16_32_BER}, we observe that an RIS-based scheme achieves significantly better SEP performance compared with the classical BPSK scheme operating in an AWGN channel. In other words, an RIS is capable of  converting a hostile wireless environment into a reliable communication channel that provides one with a low SEP for low values of the SNR, which outperforms AWGN channels. Let us analyze this behavior in detail.

By observing Fig. \ref{fig:N_16_32_BER}, we note that the average SEP has a waterfall region and a slowly-decaying region. For $\frac{N E_s}{N_0} \ll 10 $, in particular, \eqref{SEP_1_UB} unveils that $P_e$ is proportional to:
\begin{equation} \label{Approx_1}
P_e \propto \exp \left( -\frac{N^2 \pi^2 E_s}{16 N_0 } \right). 
\end{equation} 
This explains the superior performance of an RIS-based scheme. In this range, notably, even though the SNR $E_s/N_0$ is relatively low, the average SEP is quite low. This is due to the $N^2$ term inside the exponential function. 

If $\frac{N E_s}{N_0} \gg 1 $, on the other hand, \eqref{SEP_1_UB} can be approximated as follows\footnote{\label{note1}It is worth noting that $NE_s/N_0$ is measured in linear scale and the given two critical points have been determined based on several calculations for the considered values of $N$.}:
\begin{align}
	P_e & \propto \left(\frac{N (16-\pi^2)E_s}{8 N_0} \right)^{-\frac{1}{2}}  \exp \left( -\frac{N \pi^2 }{2(16- \pi^2) } \right) 
\end{align}
which explains the slowly-decaying behavior of the SEP for high values of the SNR. In particular this slowly-decaying region is caused by the scaling factor in front of the exponential function, which decays with the negative square root of the SNR. However, the SEP still decays exponentially as a function of $N$, and, therefore, $P_e$ can be reduced by appropriately increasing $N$.

\begin{figure}[!t]
	\begin{center}
		\includegraphics[width=1\columnwidth]{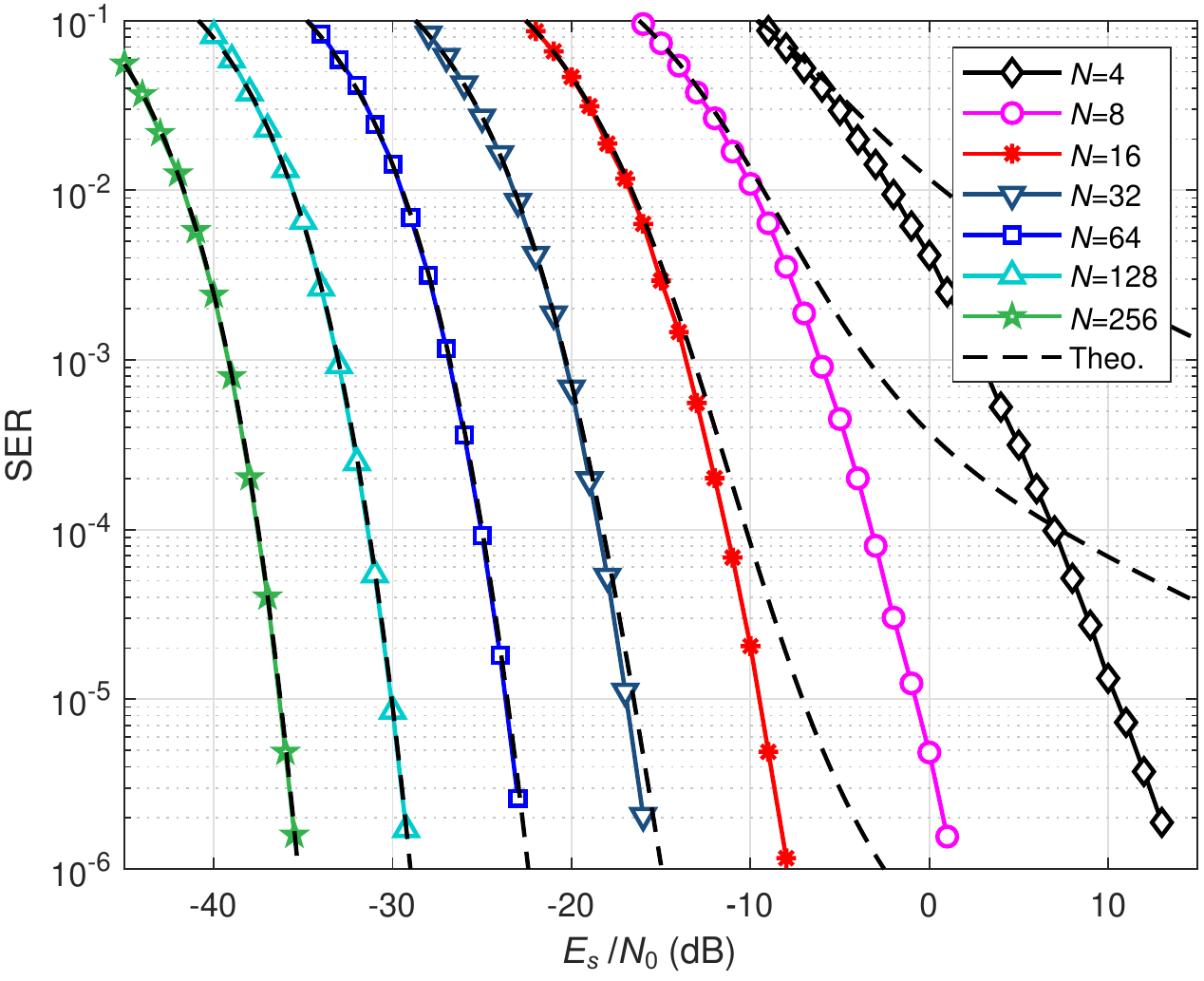}
		\caption{Simulated average SER performance (BPSK) of an RIS-based scheme with a varying number of reflecting meta-surfaces - Comparison with the theoretical formula in \eqref{SEP_1}.}
		\label{fig:Sim1}
	\end{center}
\end{figure}

In Fig. \ref{fig:Sim1}, we depict the average SEP of an RIS-based scheme for different numbers of reflecting meta-surfaces $N$ and by assuming BPSK signaling. Fig. \ref{fig:Sim1} confirms that our theoretical approximation in \eqref{SEP_1} that is based on the CLT is sufficiently accurate for large values of $N$. In the considered setup, our proposed approximation is accurate for $N \ge 32$. In particular, we note that doubling $N$ yields approximately $6$ dB improvement (four-fold decrease) of the required SNR in the waterfall region. This can be readily verified  from \eqref{Approx_1}.

Wit the aid of the MGF of the received SNR, $M_{\gamma}(s)$, we can obtain the average SEP for square $M$-QAM constellations as well \cite{Simon}:
\begin{align}\label{M_QAM}
	P_e = \,\, & \, \frac{4}{\pi} \left(1-\frac{1}{\sqrt{M}} \right) \int_{0}^{\pi/2} M_{\gamma} \left(  \frac{-3}{2(M-1) \sin^2 \! \eta}\right) d\eta  \nonumber \\
	&\hspace*{-0.7cm}- \frac{4}{\pi} \left(1-\frac{1}{\sqrt{M}} \right)^2 \int_{0}^{\pi/4} M_{\gamma} \left(  \frac{-3}{2(M-1) \sin^2 \! \eta}\right) d \eta.  
\end{align}
This integral can be upper bounded by letting $\eta=\pi/2$ and $\eta=\pi/4$ in its first and second terms, respectively. By assuming $\frac{N E_s}{N_0} \ll 10 $,  which is the SNR region of interest, the average SEP can be expressed as follows:
\begin{equation}\label{Eq:10}
P_e \propto  \exp \left( -\frac{3N^2 \pi^2 E_s}{32(M-1) N_0}\right)  
\end{equation}
where the second exponential term coming from \eqref{M_QAM} is ignored since its contribution is negligible. Since $M$ appears in the exponent of  \eqref{Eq:10}, an RIS-based scheme experiences a degradation of the error performance if the modulation order is increased, as in conventional modulation schemes. However, RISs can take advantage of large values of  $N^2$ to counteract the detrimental effect of increasing the modulation order. This can be a remedy for increasing the energy efficiency of future extreme mobile broadband applications of 6G and beyond, which will rely on high-order constellations to support extremely high data rates.

\section{Reconfigurable Intelligent Surface as a Low-Complexity and Energy-Efficient Transmitter}
In this section, we discuss the potential of using RISs as a technology that enables low-complexity and energy-efficient implementations of MIMO transmitters. The basic idea consists of illuminating an RIS with a feeder (antenna), and of encoding the data to transmit onto the phases of the signals reflected from the different reconfigurable meta-surfaces that realize the RIS. If the RIS is made of $N$ reconfigurable meta-surfaces whose reflection phase can be optimized independently of the others, then a $N$-stream virtual MIMO system can be realized by using a single RF active chain \cite{Basar_2019_LIS}. A similar solution (even though not based on reconfigurable meta-surfaces) is the concept of symbiotic radio, where a backscatter device modulates its own information over an incident signal from a transmitter by varying its reflection coefficient \cite{Long_2018}. This solution is similar to distributed SM applied to relay-aided systems \cite{DSM}.

Recently, the idea of using an RIS as a transmitter was validated with the aid of a testbed platform. In \cite{Tang_2019}, in particular, the authors have realized a $8$-PSK transmitter that utilizes a programmable surface with $256$ reconfigurable elements. By changing the bias voltage of varactor diodes, a high phase modulation resolution is obtained. The authors showed that an unmodulated carrier can be modulated by the reconfigurable meta-surface through a series of DACs that control bias voltages. In \cite{Yan_2019}, the same authors have realized a virtual quadrature phase shift keying (QPSK) constellation based on the same principle and by using a smaller reconfigurable meta-surface that is made of $128$ reconfigurable elements. In \cite{Yan_2019_2}, the idea of joint passive beamforming and data transmission is considered in the context of an RIS-assisted uplink transmission scheme. In this scenario, the authors considered the communication of a multi-antenna base station (BS) with a single-antenna user, where the on/off states of the RIS elements convey additional data. These results substantiate the potential of RISs for realizing low-complexity MIMO transmitters with a large number of equivalent radiating elements, but a few, even a single, RF chain.

\begin{figure}[!t]
	\begin{center}
		\includegraphics[width=0.9\columnwidth]{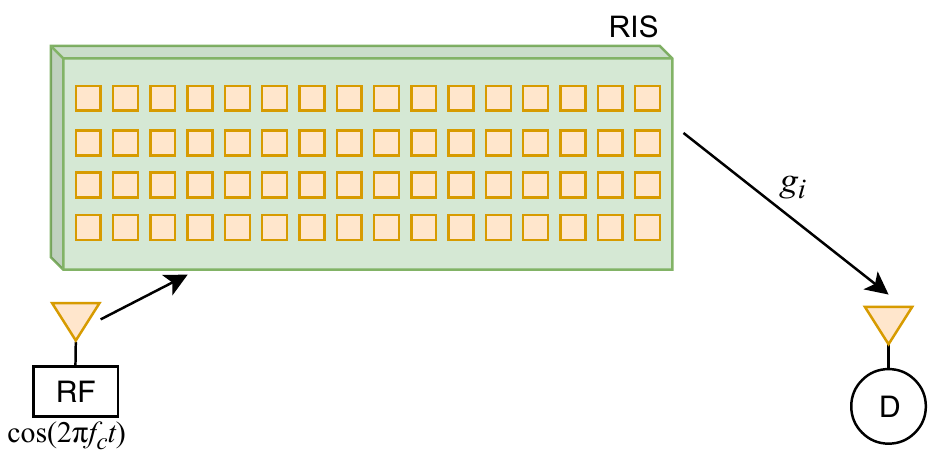}
		\caption{Using an RIS as a transmitter.}
		\label{fig:LIS_AP}
	\end{center}
\end{figure}

In this section, we analyze the error performance of RISs when utilized as transmitters. For simplicity, as single-stream transmitter is considered. The block diagram of the considered RIS-based concept is shown in Fig. \ref{fig:LIS_AP}. The RIS is illuminated by a nearby RF signal generator or contains an attachment that transmits an unmodulated carrier signal $\cos(2\pi f_c t)$ at a certain carrier frequency $f_c$ towards the RIS. The unmodulated carrier can be generated by an RF DAC with an internal memory and a power amplifier \cite{Mesleh_2018}, and information bits are conveyed only through the reflection-induced phases of the RIS. We assume that the RF source is close enough to the  RIS that its transmission is not affected by fading. On the other hand, the channel between the $i$th reflector of the RIS and D is modeled as $g_i = \beta_i e^{-j \psi_i}$.

In the considered communication scenario, RIS-induced phases carry information in addition to perform intelligent reflections that improve the received SNR. In other words, the RIS adjusts the phases of its reflecting elements with the aim of optimizing the reflected phases that maximize the received SNR, and, at the same time, that properly align the reflected signals so as to create a virtual bi-dimensional $M$-ary signal constellation diagram. For this system model, the received signal is expressed as follows:
\begin{equation}
r=\sqrt{E_s}\left[ \sum_{i=1}^N g_i e^{j \phi_i}  \right] + n 
\end{equation}
where $E_s$ is the average transmitted signal energy of the unmodulated carrier and $\phi_i$ is the reconfigurable phase induced by the $i$th reflecting element of the RIS. 

We assume that $\log_2(M)$ bits are transmitted in each signaling interval by adjusting the reflection phases so that the equality $\phi_i=\psi_i + w_m$ is fulfilled, where $w_m, m\in \left\lbrace 1,2,\ldots,M \right\rbrace $, is the message-dependent phase term introduced by the RIS that carries the information of the $m$th message. In light of this, the received signal can be expressed as follows:
\begin{equation}\label{signal}
r=\sqrt{E_s}\left[ \sum_{i=1}^N \beta_i   \right] e^{j w_m} + n=\sqrt{E_s} B e^{j w_m} + n .
\end{equation}

It is worth noting that the equivalent received signal resembles a PSK-modulated signal that is transmitted through a channel whose gain is $B$. Consequently, to minimize the average SEP, the information phases $w_1,w_2,$ $\ldots, w_M$ of this $M$-ary signaling scheme can be selected as in the classical $M$-PSK scheme, i.e., $w_m=2\pi(m-1)/M$ for $m=1,2,\ldots,M$.


Therefore, the instantaneous received SNR can be written as follows:
\begin{equation}\label{key}
\gamma = \frac{E_s B^2}{N_0}.
\end{equation}

By using again the CLT for large values of $N$ and by assuming that $\beta_i$ is a Rayleigh-distributed random variable with mean $\sqrt{\pi}/2$ and variance $(4-\pi)/4$, we obtain $B\sim \mathcal{N}(m_B,\sigma_B^2)$, where $m_B=N\sqrt{\pi}/2$ and $\sigma_B^2=N(4-\pi)/4$. Consequently, the MGF of $\gamma$, which is a non-central chi-square random variable, is the following:
\begin{equation}\label{MGF_2}
M_{\gamma}(s)= \left( \dfrac{1}{1-\frac{sN(4-\pi)E_s}{2N_0}}\right) ^{ \! \frac{1}{2}}  \! \!\exp\left( \dfrac{ \frac{sN^2 \pi E_s}{4 N_0}}{1-\frac{sN(4-\pi)E_s}{2N_0}} \right). 
\end{equation}

The  average SEP of the proposed scheme can be calculated by substituting the obtained MGF in the SEP formula of $M$-PSK signaling given in \eqref{5}. In particular, for binary signaling $(w_1=0$ and $w_2=\pi)$, we obtain the following:
\begin{equation}\label{SEP_2}
P_e= \frac{1}{\pi}  \! \int_{0}^{\pi/2} \! \! \left( \dfrac{1}{1+\frac{N(4-\pi)E_s}{2 \sin^2 \! \eta N_0}}\right) ^{\!\frac{1}{2}}  \! \!\! \exp\left( \dfrac{ -\frac{N^2 \pi E_s}{4  \sin^2 \! \eta N_0}}{1+\frac{N(4-\pi)E_s}{2 \sin^2 \! \eta N_0}} \right) \! \! d\eta. 
\end{equation}

By letting $\eta=\pi/2$ and considering the SNR range of interest $\frac{N E_s}{N_0} \ll 10 $, $P_e$ becomes proportional to:
\begin{equation} \label{Approx_3}
P_e \propto \exp \left( -\frac{N^2 \pi E_s}{4 N_0 } \right). 
\end{equation}

Two main conclusions can be drawn from \eqref{Approx_3}. If the RIS is used as a transmitter, it can convey information with high reliability, similar to using the RIS as a reflector (see Fig. \ref{fig:LIS_DH}). In addition, by comparing \eqref{Approx_1} and \eqref{Approx_3}, $1$ dB gain in the SNR is obtained by an RIS used as a transmitter with respect to using it as a reflector.

If $M$-ary signaling is considered, we can plug \eqref{MGF_2} in \eqref{5} and obtain the average SEP in the form of the definite integral as follows:
\begin{align}\label{SEP_3}
	P_e= \frac{1}{\pi} & \! \int_{0}^{(M-1)\pi/M} \! \! \left( \dfrac{1}{1+\frac{N(4-\pi) \sin^2(\pi/M)E_s}{2 \sin^2 \! \eta N_0}}\right) ^{\!\frac{1}{2}}  \nonumber 
	\\ & \times\exp\left( \dfrac{ -\frac{N^2 \pi  \sin^2(\pi/M) E_s}{4 \sin^2 \! \eta N_0 }}{1+\frac{N(4-\pi)  \sin^2(\pi/M) E_s}{2 \sin^2 \! \eta N_0}} \right) \! \! d\eta. 
\end{align}

By letting $\eta= \pi/2$, an upper-bound can be obtained. By considering the SNR range of interest, we obtain:
\begin{equation} \label{Approx_4}
P_e \propto \exp \left( - \sin^2 (\pi/M) \frac{N^2 \pi E_s}{4 N_0 } \right). 
\end{equation}

Comparing the obtained result with \eqref{Eq:10}, we conclude that an SNR loss is expected for higher order signaling $M \ge 16$. However, this loss may not be significant by taking into account that, in the relatively low SNR range of interest, the SEP can be reduced by increasing $N$.

\section{Historical Perspective and State-of-the-Art Solutions}

Within the context of unconventional wireless communication paradigms, there has been a growing interest in controlling the propagation environment in order to increase the QoS and/or spectral efficiency. IM-based  emerging schemes such as MBM \cite{Khandani_conf1,MBM_TVT,Basar_2019}, spatial scattering modulation \cite{SSM}, and beam IM \cite{BIM}, use the variations in the signatures of received signals by exploiting reconfigurable antennas or scatterers to transmit additional information bits in rich scattering environments \cite{Basar_2017}. Similarly, SM schemes \cite{SM_jour,SM_magazine,SM_magazine_2,IM_5G,Basar_2017} create a new and extended signal constellation by exploiting the indices of the available transmit antennas of MIMO systems, still benefiting from the distinguishable propagation characteristics of different transmit antennas in rich scattering environments. On the other hand, the RISs are smart devices that intentionally control the propagation environment to boost the signal quality at the receiver. In this section, we present a comprehensive survey of state-of-the-art solutions based on intelligent surfaces.

\subsection{Origins}

In this subsection, we briefly review the first concepts and ideas on reconfigurable wireless systems and intelligent surfaces that focus on the manipulation of EM waves in a deliberate manner to improve specified key performance indicators.

Transmission through intelligent walls is reported in \cite{Subrt_2012} and is based on active frequency-selective surfaces that control the signal strength. In particular, the authors focused their attention on controlling the propagation environment by influencing radio channel parameters in order to enhance the overall system performance, including signal coverage and interference. The main motivation of the authors is to change the EM properties of the material of the walls deploying frequency-selective surfaces in an indoor environment. It is shown that an active frequency-selective surface can be realized by changing the on/off states of PIN diodes, where a fully transparent or a highly reflective surface is obtained. It is shown by computer simulations that an active wall can be an efficient instrument for controlling the signal coverage and QoS.

In \cite{Subrt_2012_2}, the same authors of \cite{Subrt_2012} extended their intelligent walls concept by considering an artificial neural network (ANN) to learn the best configuration for the intelligent walls. According to sensory data and relative system performance measures, two binary outputs are obtained (walls with on and off states) by the ANN. In other words, a cognitive engine equipped with machine learning algorithms is deployed  to make decisions and control the intelligent walls. In \cite{Elzwawi_2018}, an active frequency-selective surface-based antenna is considered to enable switched beamforming by adjusting the on/off states of the PIN diodes mounted on surface panels. By exploiting the on and off combinations of the diodes on each surface panel, beam steering is achieved in various directions.

In \cite{MDR-XX}, the authors demonstrated that spatial microwave modulators are capable of shaping, in a passive way, complex microwave fields in complex wireless environments, by using only binary phase state tunable metasurfaces.

In an attempt to control the indoor wireless coverage, 3-D reflectors were realized by the authors of \cite{Xiong_2017} in order to be mounted around wireless APs. Optimized but non-reconfigurable (static-shaped) reflector designs are introduced to extend the wireless coverage in indoor environments. 

The concept of coding meta-materials is proposed in \cite{Cui_2014} for manipulating EM waves by changing the phase response of the surface elements. A meta-surface composed of binary elements (with $0$ or $\pi$ phase differences) is considered. By changing the coding bit sequences, the authors demonstrated the feasibility of changing the scattering patterns. Particularly, the authors determined the optimal coding sequences (the pattern of binary phases) to minimize the radar cross section of surfaces by redirecting the incoming EM energy into all directions. Similar to \cite{Subrt_2012}, the authors used the on/off status of PIN diodes to change the phase responses of meta-material elements. Finally, the authors constructed an FPGA hardware to control programmable meta-surfaces through PIN diodes. It is argued by the authors that these programmable meta-materials can be used to reduce the scattering features of targets and to manipulate radiation beams of antennas. The same authors introduced the concept of space-time coding digital metasurfaces in \cite{Zhang_2018} by exploiting the time dimension, which enables the control of EM waves in both space and frequency domains. In particular, the authors aimed to control not only the propagation direction but also the harmonic power distribution.

Another programmable meta-surface design that is based on PIN diodes, is proposed in \cite{Yang_2016}. It is shown by the authors that besides a reconfigurable phase,  polarization control, scattering, and beam focusing are possible through adjusting the on/off status of PIN diodes.

With the aid of reconfigurable reflect-arrays realized through tunable (varactor-tuned) resonators \cite{Hum_2014}, the concept of communications that employ smart reflect-arrays with passive reflector elements is proposed in \cite{Tan_2016}. The authors use smart reflect-arrays as an alternative to beamforming techniques that require large number of antennas to focus the transmitted or received signals, which is relatively difficult to implement for portable, wearable, or even smaller devices. It is demonstrated that reflect-arrays can be used in an effective way to change the phase of incoming signals via smart reflections without buffering or processing them, and that the received signal quality can be enhanced by optimizing the phase shift of each reflecting element of the reflect-array. Through experiments and simulations, the authors showed that  higher spectral efficiencies can be obtained by using smart
reflect-arrays without any major modification in the hardware and
software of user equipment. The authors introduced a phase optimization algorithm to maximize the transport capacity as well. In their later study, the authors investigated the feasibility of using reconfigurable reflect-arrays to support millimeter-wave communications in the context of the IEEE 802.11ad standard \cite{Tan_2018}. In this work, the authors considered electronically-controlled relay switches with on/off states in order to control reflector units instead of PIN diodes. A relatively large reflect-array was built and a beam-searching protocol was designed for application to 802.11ad networks and to overcome the LOS blockage problem of millimeter-wave networks.

As a beyond massive MIMO solution, the large intelligent surface (LIS) concept is proposed in \cite{Hu_2017,Hu_2018,Hu_2018_2}\footnote{It is worth noting that the term of LIS is also used by other researchers \cite{Basar_2019_LIS,Basar_2019_LIS_2,Alouini_2019,He_2019,Han_2018,Taha_2019,Huang_2018_2,Huang_2019} in lieu of smart reflect-arrays and RISs.}. In particular, a system with a massive number of radiating and sensing
elements, which create a contiguous and electromagnetically active surface, is considered.  A major difference between traditional massive MIMO \cite{Ericsson} and the LIS is the employment of the whole contiguous
surface for transmission and reception. The authors also introduced a signal model by considering a procedure called match filtering (through 2-D integration of the received signals on the surface) and investigated the capacity of the system.

Finally, the concept of software-controlled HyperSurfaces is proposed in \cite{Akyildiz_2018,Akyildiz_2018_2,Akyildiz_2019,Akyildiz_2018_3} with the aim of enabling full EM manipulation of the radio waves. Equipped with a lightweight Internet-of-Things (IoT) gateway, intelligent surfaces consisting of ultra-thin meta atoms receive commands from a centralized controller and adjust their EM behavior by absorbing, focusing, and steering the impinging EM waves in any direction. The functional and physical architecture of HyperSurfaces, equipped with controllable switch elements (with on/off states) and square patches, is defined. By computer simulations, it is demonstrated that the signal coverage and signal strength can be improved in indoor environments for communication at $60$ GHz, which is usually affected by high propagation losses in non-LOS transmission.

The same authors generalized their programmable wireless environment concept considering a general multi-user scenario and proposed solutions for interference minimization, eavesdropping, and multipath mitigation \cite{Akyildiz_2019}. A general optimization problem was formulated to maximize the received signal power and to minimize the maximum delay spread for arbitrarily distributed users.

\begin{table*}[t]
	{\scriptsize
		\setlength\extrarowheight{3pt}
		\caption{Initial Examples of Communications Through Intelligent Surfaces}
		\begin{tabular}{lllll}
			\textit{	Year} & \textit{Refs. }                                                & \textit{Scheme}                        & \textit{Architecture  }                                                                    & \textit{Functionality    }                                     \\ \hline\hline
			2012 & \cite{Subrt_2012,Subrt_2012_2}    & Intelligent wall              & Active frequency selective surfaces with PIN diodes                            & Fully transparent/reflecting surfaces              \\\hline
			2014 & \cite{MDR-XX} & Spatial microwave modulators &  Binary phase state tunable meta-surfaces &  Shaping complex microwave fields\\\hline
			2014 & \cite{Cui_2014}                     & Coding meta-materials          & Meta-surfaces with binary elements ($0$ or $\pi$ phases) & Reconfigurable scattering patterns                   \\\hline
			2016 & \cite{Yang_2016}                    & Programmable meta-surface      & Meta-surfaces with PIN diode-equipped cells                                     & \begin{tabular}[c]{@{}l@{}}Reconfigurable phase, polarization, \vspace*{-0.1cm}\\ and scattering\end{tabular}   \\\hline
			2016 & \cite{Tan_2016}                     & Reconfigurable reflect-arrays & Reflect-arrays with tunable (varactor-tuned) resonators                        & Adjustable reflection phase                          \\\hline
			2017 & \cite{Hu_2017,Hu_2018,Hu_2018_2} & Large intelligent surface     & Active contiguous surface for transmission and reception                        & Gains compared to  massive MIMO \\\hline
			2018 & \cite{Akyildiz_2018,Akyildiz_2019} & Software-controlled hypersurface & Meta-surfaces equipped with IoT gateways & \begin{tabular}[c]{@{}l@{}}Wave absorption, polarization, \vspace*{-0.1cm}\\ and steering\end{tabular}
			\\\hline\hline
	\end{tabular} }
\end{table*}

In Table 1, we summarize the main studies on intelligent surfaces reviewed in this subsection by highlighting their architectures and functionalities.

\subsection{State-of-The-Art Solutions}

In the last few months, several studies and innovative solutions related to RISs have been conducted by many authors. In particular, different authors have used different terms to denote the RISs, which include: \textit{reconfigurable intelligent surfaces, large intelligent surfaces, smart reflect-arrays, intelligent reflecting surfaces, passive intelligent mirrors, artificial radio space, and so on}. Specifically, researchers focused on theoretical SNR and SEP derivations, channel estimation, signal-to-interference-ratio (SINR) maximization, and joint active and passive beamforming optimization problems, investigated the application of machine learning tools, explored physical layer layer security solutions, and evaluated the potential of intelligent surfaces for application to millimeter-wave/terahertz, free space optical, and visible light communication systems. Furthermore, the first attempts on combining RISs with orthogonal frequency division multiplexing (OFDM) and SM/space shift keying (SSK) schemes have been reported. 

In this section, we briefly describe research works on intelligent surfaces, which appeared in the literature during the past 1-2 years by following a historical order.

\begin{figure}[!t]
	\begin{center}
		\includegraphics[width=0.9\columnwidth]{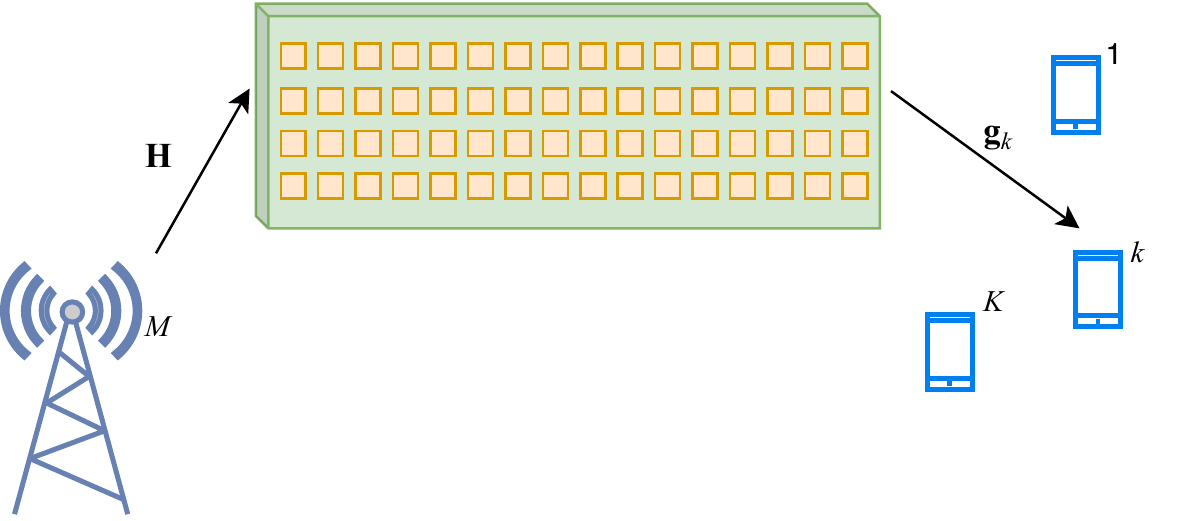}
		\caption{Multi-user downlink transmission through an RIS (the LOS path is optional).}
		\label{PIM}
	\end{center}
\end{figure}

An RIS-assisted downlink transmission scenario is considered in \cite{Huang_2018} to support multiple users. More specifically, the authors considered the system model shown in Fig. \ref{PIM}, where a multi-antenna BS supports $K$ single-antenna users through intelligent reflections without a clear LOS path between the users and the BS. The RIS elements only reflect phase shifted versions of the incoming signals as in the model of Fig. \ref{fig:LIS_DH}. As a result, denoting the number of BS antennas and RIS elements (reflecting meta-surfaces) by $M$ and $N$, respectively, the received signal at user $k$ can be expressed as
\begin{equation}\label{users}
r_k= \mathbf{g}_k^{\mathrm{T}}\Phi \mathbf{H} \mathbf{x} + n_k
\end{equation}
where $\mathbf{x} \in \mathbb{C}^{M \times 1}$ is the precoded data vector, $\mathbf{H} \in \mathbb{C}^{N \times M}$ is the matrix of channel coefficients of the BS-RIS link, $\mathbf{g}_k \in \mathbb{C}^{N \times 1} $ is the vector of channel coefficients between the RIS and user $k$, and $n_k$ is the AWGN sample at this user. Similar to \eqref{eq:matrix}, $\Phi=\mathrm{diag} ( \left[ e^{j\phi_1} \,\, e^{j\phi_2}\,\,\cdots\,\,e^{j\phi_N} \right] )$ is a diagonal matrix that yields the phase shifts applied by the reflecting elements of the RIS. The authors focused their attention on the maximization of the system sum-rate by optimizing the phases of the RIS and user powers. The resulting non-convex optimization problem is solved by combining  alternating maximization and majorization-minimization algorithms, and  improvements  in the overall system throughput are reported.

In \cite{Huang_2018_2}, the same authors studied the maximization of the energy efficiency by  considering RIS reflectors with finite resolution phases. It is shown by computer simulations that even $ 1 $-bit phase ($0$ and $\pi$) resolution reflecting elements  increase the energy efficiency of the system  compared  to conventional AF relaying systems. In \cite{Huang_2019}, the authors reported more comprehensive simulation results on energy efficiency and system sum-rate for a more practical system setup and system parameters. It is worth noting that the implementation of this scheme requires full channel state information (CSI) and the knowledge of RIS phase terms at the BS for transmit beamforming.

Th problem of joint active and passive beamformer design is investigated in \cite{Wu_2018} and \cite{Wu_2018_3}. The authors, in particular, analyzed the minimization of the total transmit power at the BS by jointly optimizing the transmit beamforming vectors of the BS and the phase shifts of the RIS by imposing SINR constraints for the users considering a multi-user downlink communication system. The obtained non-convex optimization problem is tackled by employing semidefinite relaxation and alternating optimization techniques. The authors showed that the transmit power of the BS scales by $1/N^2$ as the number of reflector units $N$ goes to infinity. In their follow-up study \cite{Wu_2018_2}, the authors considered the same power minimization problem by assuming discrete (finite resolution) RIS phase shifts. It is shown that, compared to the ideal case with continuous phase shifts, an RIS with discrete phase shifts achieves the same power scaling law as a function of $ N $ as for infinite-resolution phase shifts but a constant performance loss is observed. The effect of erroneous reflector phases on the error performance is investigated recently as well \cite{Coon_2019}.

In \cite{Basar_2019_LIS}, the author introduced a mathematical framework  for the calculation of the average SEP of RIS-assisted systems. Furthermore, the possibility of using an RIS as an AP (transmitter) by utilizing an unmodulated carrier for intelligent reflection is studied. With the aid of computer simulations, it is shown that doubling the number of reflecting elements provides one with a $6$ dB reduction of the SNR to obtain the same SEP. In other words, the average received SNR of RIS-based systems is shown to be proportional to the square of the total number of reflecting elements of the RIS $(N)$. In the follow-up study \cite{Basar_2019_LIS_2}, the author considered the concept of intelligent surfaces-assisted IM by proposing LIS-SM and LIS-SSK schemes. In these schemes, the optimization of the phases of intelligent surfaces and the concept of IM are jointly exploited for improving the signal quality and the spectral efficiency. Maximum energy-based suboptimal (greedy)
and exhaustive search-based optimal (maximum likelihood) detectors are formulated, and a unified
framework to calculate the theoretical
average bit error probability is proposed. It is shown by computer simulations that the resulting bit error probability is significantly better than conventional
fully-digital precoding-based receive SSK schemes. A trade-off between the receiver cost and the BER performance is investigated as well.

The authors of \cite{Alouini_2019} investigated the maximization of the minimum SINR for an RIS-assisted multi-user MISO scheme in rank-one or full-rank LOS channels between the BS and the RIS. Furthermore, correlated Rayleigh channels are considered between RIS elements and users, and an algorithm to optimize the RIS phases in the presence of large-scale fading is proposed. For the same system, the authors of \cite{Schober_2019} focused their attention on optimizing the transmit beamformer and RIS phases, and proposed two  algorithms to jointly optimize the beamformer at the AP and the phase shifts at the RIS by considering the maximum achievable spectral efficiency. It is shown that, unlike \cite{Wu_2018} and \cite{Alouini_2019}, the proposed algorithms guarantee locally optimal solutions with the perfect knowledge of channels at the AP and the RIS.

In \cite{Han_2018}, the authors considered an RIS-assisted large-scale antenna system in which a BS communicates with a single-antenna user. Similar to previous works, assuming the knowledge of CSI and RIS phases at the BS, precoding can be performed, and the ergodic capacity of the system is maximized through the optimization of the RIS phases. The authors also investigated the effect of different Rician $K$ factors and discrete phase shifts. It is shown by computer simulations that the use of $ 2 $-bit phase shifts can ensure a high capacity.

The authors of \cite{Mishra_2019} proposed a new channel estimation protocol for an RIS-assisted single-user MISO system with energy harvesting. In particular, the authors take into account that RISs do not have active components. To enable efficient power transfer, both active and passive near-optimal beamforming designs are formulated. In \cite{Taha_2019}, with the motivation of reducing the training overhead in the presence of passive RIS elements, the authors considered two separate methods for the RIS design in the presence of unknown channel knowledge: one is based on compressive sensing and the other is based on deep learning. In this work, a novel RIS architecture based on sparse channel sensors, in which some of the  RIS units are active (have a connection to a baseband processor), is proposed. In other words, a number of active elements are used at the RIS to simplify the channel estimation process. The authors  considered a deep learning-based solution, in which the RIS learns how to interact with the incoming signals in an optimal way by exploiting the estimated channels at the active elements. On the other hand, the authors of \cite{He_2019} considered the problem of cascaded channel estimation with fully passive RIS elements for the communication of multi-antenna terminals through an RIS. A general concept is proposed for the estimation of the S-RIS-D cascaded MIMO channel. 

For the scenario of multi-user uplink transmission, the performance of the over-the-air computation scheme, in which the AP computes a target function of the aggregated data from all users, is investigated in \cite{Jiang_2019} by exploiting RISs.

The physical layer security of RIS-based systems is investigated in \cite{Schober_2019_2} and \cite{Chen_2019}. Specifically, the authors of \cite{Schober_2019_2} considered an RIS-assisted secure communication system that consists of a legitimate receiver and an eavesdropper. Both the beamformer at the BS and the RIS phase shifts are jointly optimized to enhance the secrecy rate. Furthermore, the secrecy rate of massive MIMO and RIS-aided systems is compared and it is revealed that increasing the number of RIS reflecting elements is more beneficial than increasing the number of antenna elements at the BS. A downlink MISO broadcast system with multiple legitimate receivers and eavesdroppers is considered in \cite{Chen_2019}. The authors formulated a minimum-secrecy-rate maximization problem by jointly optimizing the beamformer at the BS and reflecting coefficients (both discrete and continuous) at the RIS. Globally optimal and low-complexity suboptimal algorithms are proposed. Recently, the authors of \cite{Cui_2019} focused on the problem of stronger eavesdropping channel and maximized the secrecy rate of
the legitimate user by jointly designing the AP's transmit beamformer and the RIS phases.

Recent research contributions on the design of HyperSurfaces include the following. A sensing system to estimate the radio waves impinging upon the HyperSurfaces
that can operate without additional external or internal hardware such as field nano-sensors, is proposed in \cite{Liaskos_2019}. It is shown that HyperSurfaces can be programmed for wave sensing and wave manipulation simultaneously. Finally, an approach based on machine learning (neural networks) to adaptively configure HyperSurfaces-aided programmable environments is proposed in \cite{Liaskos_2019_2}.

As for using RISs for transmission and reception, researchers focused their attention on outage probability \cite{Jung_2019_1}, asymptotic data rate \cite{Jung_2019_2}, and uplink spectral efficiency \cite{Jung_2019_3}.

More recently, researchers considered RIS-assisted millimeter-wave/terahertz communications \cite{Nie_2019}, visible light communications \cite{Valagiannopoulos_2019}, free space optical communications \cite{Najafi_2019}, and OFDM  systems \cite{Yang_2019}.

\begin{table*}[t]
	{\scriptsize
		\setlength\extrarowheight{3pt}
		\caption{State-of-the-Art Schemes with Intelligent Surfaces}
		\label{tab:2}
		\begin{tabular}{ll}
			\textit{Refs.}                                        & \textit{Contribution}                                                            \\ \hline\hline
			\cite{Huang_2018,Huang_2018_2,Huang_2019} & \begin{tabular}[c]{@{}l@{}}	A precoding-aided RIS scheme is considered for multi-user downlink transmission. \vspace*{-0.1cm}\\ Maximization of sum-rate and energy efficiency is performed with finite resolution reflectors.\end{tabular} \\\hline
			\cite{Wu_2018,Wu_2018_3,Wu_2018_2} & \begin{tabular}[c]{@{}l@{}} The problem of joint active and passive beamformer design is investigated for a MISO multi-user system.	\vspace*{-0.1cm}\\ Minimization of the BS transmit power is performed through optimization and square-law scaling in transmit power is demonstrated.  \end{tabular}
			\\\hline 
			\cite{Basar_2019_LIS}	 & \begin{tabular}[c]{@{}l@{}} A mathematical framework is proposed for the calculation of the average SEP of RIS (LIS)-assisted systems.\vspace*{-0.1cm}\\ The concept of using the RIS as an AP (transmitter) is also introduced.
			\end{tabular}
			\\\hline 
			\cite{Basar_2019_LIS_2}	 & \begin{tabular}[c]{@{}l@{}}The  concept of RIS (LIS)-assisted IM is proposed by considering the RIS as an AP.\vspace*{-0.1cm}\\ Greedy and maximum likelihood detectors are formulated for LIS-SM and LIS-SSK schemes along with theoretical derivations.
			\end{tabular}

			\\\hline 
			\cite{Alouini_2019}	 & \begin{tabular}[c]{@{}l@{}} The maximization of the minimum SINR is investigated for an RIS-assisted multi-user MISO system.\vspace*{-0.1cm}\\ Rank-one and full-rank LOS channels, correlated RIS channels, and large-scale fading statistics are considered for phase optimization.
			\end{tabular}
			
			\\\hline 
			\cite{Schober_2019}	 & \begin{tabular}[c]{@{}l@{}} The problem of optimal transmit beamformer and RIS phase shifter is investigated to maximize the achievable spectral efficiency.\vspace*{-0.1cm}\\ It has been shown that the proposed algorithms guarantee locally optimal solutions.
			\end{tabular}

			\\\hline
			\cite{Han_2018}	 & \begin{tabular}[c]{@{}l@{}} An RIS-assisted large-scale MISO system is considered with Rician fading.\vspace*{-0.1cm}\\ Ergodic capacity of the system is maximized by the optimization of LIS phases.
			\end{tabular}
			\\\hline 
			\cite{Mishra_2019}	 & \begin{tabular}[c]{@{}l@{}} A new channel estimation protocol for an RIS-assisted MISO system with energy harvesting is proposed.\vspace*{-0.1cm}\\ Active and passive near-optimal beamforming designs are formulated to enable efficient power transfer.
			\end{tabular}
			\\\hline 
			\cite{Taha_2019}	 & \begin{tabular}[c]{@{}l@{}} 
				A new RIS architecture based on sparse channel sensors, in which some of the existing RIS units are active, is proposed. \vspace*{-0.1cm}\\Two separate methods, based on compressive sensing and deep learning, are considered for the RIS design.
			\end{tabular}
			\\\hline 
			\cite{He_2019}	 & \begin{tabular}[c]{@{}l@{}} Considered the problem of cascaded channel estimation with fully passive RIS elements. \vspace*{-0.1cm}\\ An RIS-assisted massive multi-user MIMO system is considered and a three-stage channel estimation algorithm is proposed.
			\end{tabular}
			\\\hline 
			\cite{Schober_2019_2}	 & \begin{tabular}[c]{@{}l@{}} An RIS-assisted secure communication system with a legitimate receiver and an eavesdropper is considered.\vspace*{-0.1cm}\\ Showed that increasing the number of RIS reflecting elements is more beneficial than increasing the number of antenna elements at BS.
			\end{tabular}
			\\\hline 
			\cite{Chen_2019}	 & \begin{tabular}[c]{@{}l@{}} A downlink MISO broadcast system with multiple legitimate receivers and eavesdroppers is considered.\vspace*{-0.1cm}\\ A minimum-secrecy-rate maximization problem is formulated by jointly optimizing the BS beamformer and RIS reflecting coefficients.
			\end{tabular}
			\\\hline \hline
		\end{tabular} 
	}
\end{table*}

The emerging concept of programmable radio environments is receiving considerable attention from the research community of computer scientists as well. In the last couple of years, in particular, a few large scale-scale testbeds have been realized, and promising results have been obtained. Notable examples include \cite{MDR_A1,MDR_B1,MDR_C1}. The underlying idea of these research activities is to prove the feasibility of moving beamforming functions from the radio end-points to the environment. In \cite{MDR_C1}, in particular, the author realized a testbed that is today considered to be the largest RIS currently available in the world: The prototype has $ 3,720 $ inexpensive antennas that are deployed on a $ 6 $ square-meter surface. Experimental results in an indoor environment have shown an improvement of the median of the signal strength of $ 10.5$x, and an improvement of the median of the data rate of $ 2.1$x  \cite[Fig. 11]{MDR_C1}.

In Table \ref{tab:2}, we present a summary of the major contributions of the recent studies reviewed in this subsection.

\section{Potential Use Cases}

In this section, we briefly discuss a few use cases in which the RISs may play a fundamental role either for enhancing the coverage probability, spectral efficiency, and energy efficiency or for reducing the implementation complexity and power consumption of wireless networks. Five potential use cases are briefly discussed.

\textbf{Overcoming non-LOS Scenarios} -- One of the most promising use cases to leverage the RISs in wireless networks consists of employing them as reconfigurable reflectors in scenarios in which the LOS path is either blocked or is not strong enough to support cell-edge users. For example,  RISs can be easily attached to walls or ceilings in indoors, and can be integrated into the facades of buildings in outdoors. This application scenario seems to be relevant in the high frequency transmission range, e.g., in the millimeter-wave spectrum, in the D-band spectrum ($>100$ GHz), and in the visible light spectrum. In these cases, in fact, the LOS path is often obstructed, and the possibility of realizing strong and reconfigurable non-LOS links is a promising use case, especially because of the promising scaling laws as a function of the distance and of the number of reflecting meta-surfaces that are obtained in Section II. Field trials that substantiate the applicability and potential gains of this case study were recently reported for application to vehicular networks by research scientists of NTT Docomo and Metawave Corporation \cite{MDR-17}.

\textbf{Overcoming Localized Coverage Holes} -- Another promising case study consists of using the RISs to counteract localized coverage holes in urban scenarios, and indoor harsh propagation environments. In many urban and densely populated cities worldwide, in fact, there exist localized dead zones where the signal quality is not sufficiently good. Similar issues occur in indoor environments, like industrial factories and underground metro stations. In these scenarios, conventional solutions to overcome coverage holes are to deploy more BSs or relays/repeaters. Unfortunately, these solutions are expensive and increase the carbon footprint of wireless communications. The deployment of RISs is, on the other hand, a cost-effective and environmental-friendly solution to solve the problems of localized coverage holes.

\textbf{Reducing the EM Pollution} -- In contrast to other communication technologies, one of the main features of RISs is recycling the radio waves in a constructive and energy-efficient manner. Multipath propagation is, in fact, often perceived as incontrollable and is usually counteracted by increasing the complexity of the transmitters and the receivers. This usually entails an increase of the number of radio waves emitted, e.g., by deploying more BSs or relays, which produce additional signals in the environment. This results in increasing the emission of EM radiations. The use of RISs, on the other hand, does not foresee the generation of new signals, but their intelligent utilization. So, the concept of RISs is a promising solution to lower the levels of EM radiations, with major applications in scenarios like, hospitals and airplanes.

\textbf{Energy-Free Internet of Things} -- The IoT is regarded as a fundamental component of 5G and 6G  wireless networks. The possibility of collecting data from large numbers of sensors deployed throughout the network has a myriad of applications. These devices, however, are expected to communicate the sensed data to fusion centers, which are then in charge of the subsequent processing and analysis. The total amount of energy that these devices need in order to report the sensed data cannot be underestimated, and is one of the bottlenecks to make the IoT a pervasive technology. The concept of RISs in combination with backscatter communications is a promising alternative to allow IoT devices to report the sensed data in a energy free manner. Imagine to realize clothings and IoT devices with reconfigurable meta-surfaces that are capable of modifying the reflected waveforms according to the sensed data. By using the backscatter communications principle, the sensed data may be piggybacked into the shape of the reflected signals at no overhead and at a zero energy cost \cite{Di_Renzo_2019}.

\textbf{Low-Complexity and Energy-Efficient Massive Transmitters} -- The advantages of using MIMO, and, more recently, massive MIMO are undeniable. However, these gains are not obtained for free. In particular, the benefits of MIMO come at the cost of using a number of RF chains that is usually the same as the number of radiating elements. As a result, the complexity and power consumptions associated with using large numbers of antennas cannot be under estimated. The use of RISs as transmitters offers a unique opportunity to realize very large antenna arrays with a few, possibly one, RF chain \cite{Basar_2019_LIS}. The testbed recently implemented in \cite{Tang_2019} and \cite{Yan_2019_2} is a tangible example that realizing low-complexity massive MIMO is possible, by leveraging the principles of SM, MBM, and, more in general, IM.

\section{Open Research Issues}

In this section, we briefly discuss the major open research issues that need to be tacked in order to leverage the potential of RISs in wireless networks. We discuss, in particular, how the use of RISs necessitates to rethink how wireless networks are analyzed and designed today.

\textbf{Physics- and EM-Compliant Models} -- A major limitation of current research on RISs in wireless networks is the lack of accurate and tractable models that describe the reconfigurable meta-surfaces as a function of their EM properties. The vast majority of research works available to date rely, e.g., on the assumption that the meta-surfaces act as perfect reflectors. However, the meta-surfaces are meant to apply functions other than reflection, and, more importantly, their response to the radio waves depends on several factors, among which the angle of incidence, the angle of reflection, the angle of refraction, the polarization of the incident wave, and the specific materials that the meta-surface is made of. Also, the size of the meta-surface plays an important role on its properties. Physics- and EM-compliant models are, therefore, needed in order to avoid studying too simplistic system models and to obtain too optimistic performance predictions. The zero-thickness sheet model of the meta-surfaces is a good starting point to build such models \cite{MDR-20}. 

\textbf{Experimentally-Validated Channel Models} -- To date, there exist no experimentally-validated channel models that provide wireless researchers with accurate and realistic information on the path-loss, shadowing, and fast-fading statistics for RISs. In Section II, we have predicted that the path-loss function of an RIS that acts as a perfect reflector scales with the sum of the distances between the RIS and the transmitter and receiver. This theoretical model is based on geometrical optics approximations (and the application of Fermat's principle), whose accuracy depends on the geometry and the materials that the meta-surface is made of. In addition, the signals reflected from the elementary meta-surfaces or even by each of their individual scattering particles may not necessarily be uncorrelated. Some level of spatial correlations are likely to be expected, which may affect the ultimate performance limits and achievable scaling laws.

\textbf{Information- and Communication-Theoretic Models} -- The use of RISs renders the information- and communication-theoretic models employed in wireless communications obsolete \cite[Fig. 6]{Di_Renzo_2019}. The conventional formulation of the Shannon capacity, in particular, needs to be revisited, since the system itself can be programmed, and the distribution of the input needs to be adapted to the possible states that the system itself can assume. The possible states of the system are dictated by the functions that are applied by the meta-surfaces, and by their statistical distribution. The achievable limits of the RISs and their interpretation in small-scale and large-scale wireless networks need to be determined. In the presence of RISs, in fact, not only the input is an optimization variable, but the system itself becomes an optimization variable, and they need to be jointly optimized.

\textbf{Communication-Theoretic Performance Limits} -- With the exception of \cite{Basar_2019_LIS,Basar_2019_LIS_2,MDR-21}, the optimization of, e.g., the phases of the RISs that act as intelligent reflecting surfaces are obtained by solving complicated, and often non-convex, optimization problems. The optimal reflecting phases, which are channel-dependent, are, therefore, only available numerically. This prevents communication theorists from developing tractable analytical frameworks for analyzing the performance of RIS-empowered wireless networks, and to shed light on the corresponding performance trends. In the present paper, we have proposed simple analytical frameworks under simple system models, where the optimal phases can be easily determined. By slightly modifying the system model as in \cite{MDR-21}, we readily note that the analytical complexity of the resulting optimal phases to employ increases significantly. Due to the insights for system design that we gained in the previous sections based on  simple system models, it is considered to be important to develop tractable and insightful communication-theoretic frameworks for unveiling the achievable performance of RISs in future wireless networks, as a function of their many constituent parameters.

\textbf{Spatial Models for System-Level Analysis and Optimization} -- As just mentioned, the analysis and optimization of point-to-point wireless networks in the presence of RISs is an open and not an easy task. More difficult is, therefore, the analysis and optimization of large-scale networks where multiple RISs are spatially distributed according to some complex spatial patterns. A simple framework was recently introduced in \cite{MDR-22} by using the theory of line processes, and under the assumption that the RISs can reflect signals coming from any direction and can direct them towards any direction. In contrast with typical approaches that are used in the literature for system-level analysis, which are primarily based on stochastic geometry models, the use of RISs necessitates new models and methods of analysis. It is known, in fact, that signals' reflections are usually ignored when analyzing large-scale networks due to the inherent analytical complexity \cite{MDR-23}. Furthermore, the spatial correlation that originates from the spatial positions of objects, transmitters, and receivers is not easy to be modeled in a tractable manner \cite{MDR-24}. In RIS-empowered wireless networks, reflections, refractions, and blockages due to the objects are exploited and optimized for increasing the network performance, since the RISs aim to take advantage of them. Major open issues on the modeling and analysis of large-scale networks need, therefore, to be solved.

\textbf{Channel Estimation and Feedback Overhead} -- One of the fundamental assumptions that make RIS-assisted communications a competitive low-complexity and energy-efficient technology lies in its almost passive implementation. This implies that, ideally, the RISs are expected not to rely on power amplifiers, channel estimators, ADCs, DACs, etc. Therefore, the fundamental issue lies in how the RISs can estimate the channels that are needed for optimizing, e.g., the reflection phases, and how they can report them to the transmitter, the receiver, or any network controller in charge of computing the optimal phases as a function of the channels themselves. A possible solution may be to embed low-power sensors throughout the RISs, possibly powered by energy harvested modules, which are in charge of sensing the channels, and then reporting them to a gateway that is located on the RISs, which, in turn, sends them to the network controller. This solution relies on the availability of at least one power amplifier and one ADC, but has the advantage that no computation is made on the RISs \cite{Di_Renzo_2019}. Another promising solution relies on estimating the optimal phases without assuming the knowledge of the individual channels between the RIS and the transmitter and the receiver, but by considering only the combined (or product)  channel between them. This solution was recently reported in \cite{He_2019}. The optimality of this approach compared with knowing the individual channels, and the associated complexity vs. performance vs. estimation overhead trade-offs are an important open issue. More in general, the design of efficient channel estimation and feedback protocols that make the RISs as passive as possible, and the associated performance vs. complexity trade-offs are not known yet.

\textbf{Implementation, Testbeds, and Field Trials} -- In the last few years, a few promising testbeds and experimental activities to substantiate the potential gains offered by  RISs have been reported \cite{MDR-17,MDR-XX,Tan_2018,Akyildiz_2019,MDR_C1}. These research activities are promising, and have initially confirmed that the use of RISs in wireless networks is a practically viable solution. However, the details of these experimental tests are often difficult to find, and are not sufficient to judge the actual potential of RISs in realistic operating conditions. 

\textbf{Data-Driven Optimization} -- RIS-empowered wireless networks are much more complex systems to model, analyze, and design than current and next-generation wireless networks. Due to their inherent system complexity, the use of data-driven methods based on deep learning, reinforcement learning, and transfer learning are promising tools to leverage \cite{MDR-8}. Machine learning, in particular, is a promising solution to simplify the implementation of RIS-based communication systems, and to increase the efficiency of the communication system  \cite{Taha_2019,Liaskos_2019_2}.

\textbf{Integration of RISs with Emerging Technologies} -- The concept of RIS-empowered wireless networks has at its core the fundamental idea of moving the functions that are usually implemented at the radio end-points of a communication network, i.e., the base stations, the APs, and the mobile terminals, to the environment. In that regard, the synergistic integration of RISs with state-of-the-art and emerging technologies, such as small cells, massive MIMO, millimeter-wave communications, visible light communications, terahertz communication, free space optics, IoT, drones-aided communications, energy harvesting, physical layer security, beamforming, etc., is a promising and unexplored research direction. For example: i) RISs can be used for sculpting and exploiting non-LOS reflections at very high transmission frequencies, e.g., above $ 100 $ GHz, where it is very difficult to have reliable LOS links in mobile environments \cite{MDR-17}, ii) RISs can be leveraged to introduce a new concept of beamforming that is not realized at the transmitter anymore, but is moved to the environment, thus enabling low-complexity and tiny single-antenna devices to benefit from high-precision beamforming \cite{MDR_C1}, and iii) RISs may be a fundamental enabler to realize advanced security schemes at the physical layer, thanks to their potential of sculpting the wireless environment \cite{Schober_2019_2,Chen_2019,Cui_2019}.

\section{Conclusion}
In this paper, we have summarized the latest research activities on the emerging field of RIS-empowered wireless networks. We have illustrated the main differences that render RISs a new technology compared with, at the first sight, similar technologies, such as relaying and backscatter communications. We have described two possible uses to leverage the potential of RISs in wireless networks: to shape the radio waves in order to control, in a deterministic fashion, the multipath propagation, e.g., the signals reflected from walls are steered toward specified directions, and to realize low-complexity and energy efficient transmitters that require only a limited, ideally one, active RF chain. The error probability of both implementations has been studied by capitalizing on the CLT, and it has been shown that the error probability exhibits a water-fall behavior as a function of the number of reconfigurable elements of the RISs and of the SNR. As far as the link budget analysis is concerned, we have highlighted that the concept of RISs is different from relaying and backscatter communications, since, if their geometric size is sufficiently larger than the wavelength, they can be treated as specular reflectors, and the received power as a function of the distance is determined, at the first-order, by the Fermat's principle. Finally, we have illustrated possible use cases where the RIS may play a significant role, and have discussed fundamental research issues to tackle in order to fully exploit the potential of RISs in wireless networks.

\bibliographystyle{IEEEtran}
\bibliography{bib_2019}

\newpage

\begin{IEEEbiography}[{\includegraphics[width=1in,height=1.25in,clip,keepaspectratio]{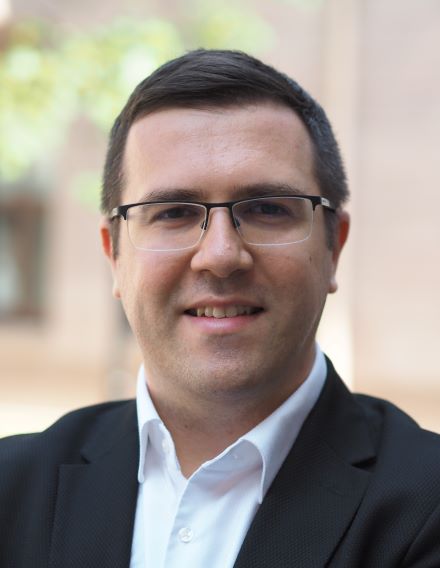}}]{Ertugrul Basar}  (S'09-M'13-SM'16) was born in Istanbul, Turkey, in 1985. He received the B.S. degree (Hons.) from Istanbul University, Turkey, in 2007, and the M.S. and Ph.D. degrees from Istanbul Technical University, Turkey, in 2009 and 2013, respectively. He is currently an Associate Professor with the Department of Electrical and Electronics Engineering, Ko\c{c} University, Istanbul, Turkey and the director of Communications Research and Innovation Laboratory (CoreLab). His primary research interests include MIMO systems, index modulation, waveform design, visible light communications, and signal processing for communications.
	
	Recent recognition of his research includes the Science Academy (Turkey) Young Scientists (BAGEP) Award in 2018, Mustafa Parlar Foundation Research Encouragement Award in 2018, Turkish Academy of Sciences Outstanding Young Scientist (TUBA-GEBIP) Award in 2017, and the first-ever IEEE Turkey Research Encouragement Award in 2017. 
	
	Dr. Basar currently serves as an Editor of the \textit{IEEE Transactions on Communications} and \textit{Physical Communication} (Elsevier), and as an Associate Editor of the \textit{IEEE Communications Letters}. He served as an Associate Editor for the \textit{IEEE Access} from 2016 to 2018. 
	
\end{IEEEbiography}

\begin{IEEEbiography}[{\includegraphics[width=1in,height=1.25in,clip,keepaspectratio]{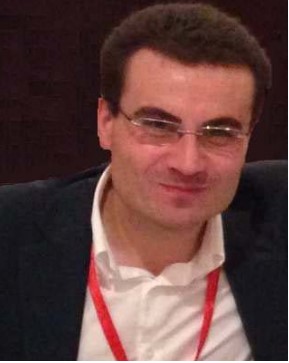}}]{Marco Di Renzo} (S'05-AM'07-M'09-SM'14) was born in L'Aquila, Italy, in 1978. He received the Laurea (cum laude) and Ph.D. degrees in electrical engineering from the University of L'Aquila, Italy, in 2003 and 2007, respectively, and the Habilitation à Diriger des Recherches (Doctor of Science) degree from University Paris-Sud, France, in 2013. Since 2010, he has been a Chargé de Recherche CNRS (CNRS Associate Professor) in the Laboratory of Signals and Systems (L2S) of Paris-Saclay University - CNRS, CentraleSupélec, Univ Paris Sud, Paris, France. He serves as the Associate Editor-in-Chief of \textit{IEEE Communications Letters}, and as an Editor of \textit{IEEE Transactions on Communications}, and \textit{IEEE Transactions on Wireless Communications}. He is a Distinguished Lecturer of the IEEE Vehicular Technology Society and IEEE Communications Society. He is a recipient of several awards, including the 2013 IEEE-COMSOC Best Young Researcher Award for Europe, Middle East and Africa, the 2013 NoE-NEWCOM Best Paper Award, the 2014-2015 Royal Academy of Engineering Distinguished Visiting Fellowship, the 2015 IEEE Jack Neubauer Memorial Best System Paper Award, the 2015-2018 CNRS Award for Excellence in Research and Ph.D. Supervision, the 2016 MSCA Global Fellowship (declined), the 2017 SEE-IEEE Alain Glavieux Award, the 2018 IEEE-COMSOC Young Professional in Academia Award, and 7 Best Paper Awards at IEEE conferences (2012 and 2014 IEEE CAMAD, 2013 IEEE VTC-Fall, 2014 IEEE ATC, 2015 IEEE ComManTel, 2017 IEEE SigTelCom, EAI 2018 INISCOM, IEEE ICC 2019).
\end{IEEEbiography}

\begin{IEEEbiography}[{\includegraphics[width=1in,height=1.25in,clip,keepaspectratio]{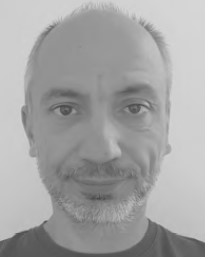}}]{Julien  De Rosny} received the M.S. and Ph.D.
	degrees in wave physics from University Pierre
	and Marie Curie, Paris, France, in 1996 and 2000,
	respectively. He held a Postdoctoral position with
	the Scripps Research Institute, CA, USA, from
	2000 to 2001. In 2001, he joined with the Laboratoire
	Ondes et Acoustique, CNRS, France.
	Since 2014, he has been a CNRS Senior Scientist
	with the Institut Langevin, Paris, France. His
	research interests include telecommunications in
	complex media, acoustic, and electromagnetic waves-based imaging.
\end{IEEEbiography}

\begin{IEEEbiography}[{\includegraphics[width=1in,height=1.25in,clip,keepaspectratio]{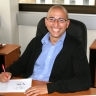}}]{Mérouane Debbah} (S'01-M'04-SM'08
	-F’15) received the M.Sc. and Ph.D. degrees from the Ecole Normale Supérieure Paris-Saclay, France. He was with Motorola Labs, Saclay, France, from 1999 to 2002, and also with the Vienna Research Center for Telecommunications, Vienna, Austria, until 2003. From 2003 to 2007, he was an Assistant Professor with the Mobile Communications Department, Institut Eurecom, Sophia Antipolis, France. From 2007 to 2014, he was the Director of the Alcatel-Lucent Chair on Flexible Radio. Since 2007, he has been a Full Professor with CentraleSupelec, Gif-sur-Yvette, France. Since 2014, he has been a Vice-President of the Huawei France Research Center and the Director of the Mathematical and Algorithmic Sciences Lab. He has managed 8 EU projects and more than 24 national and international projects. His research interests lie in fundamental mathematics, algorithms, statistics, information, and communication sciences research. He is an IEEE Fellow, a WWRF Fellow, and a Membre émérite SEE. He was a recipient of the ERC Grant MORE (Advanced Mathematical Tools for Complex Network Engineering) from 2012 to 2017. He was a recipient of the Mario Boella Award in 2005, the IEEE Glavieux Prize Award in 2011, and the Qualcomm Innovation Prize Award in 2012. He received 20 best paper awards, among which the 2007 IEEE GLOBECOM Best Paper Award, the Wi-Opt 2009 Best Paper Award, the 2010 Newcom++ Best Paper Award, the WUN CogCom Best Paper 2012 and 2013 Award, the 2014 WCNC Best Paper Award, the 2015 ICC Best Paper Award, the 2015 IEEE Communications Society Leonard G. Abraham Prize, the 2015 IEEE Communications Society Fred W. Ellersick Prize, the 2016 IEEE Communications Society Best Tutorial Paper Award, the 2016 European Wireless Best Paper Award, the 2017 Eurasip Best Paper Award, the 2018 IEEE Marconi Prize Paper Award, the 2019 IEEE Communications Society Young Author Best Paper Award and the Valuetools 2007, Valuetools 2008, CrownCom 2009, Valuetools 2012, SAM 2014, and 2017 IEEE Sweden VT-COM-IT Joint Chapter best student paper awards. He is an Associate Editor-in-Chief of the journal Random Matrix: Theory and Applications. He was an Associate Area Editor and Senior Area Editor of the \textit{IEEE Transactions on Signal Processing} from 2011 to 2013 and from 2013 to 2014, respectively. He is the co-founder of Ximinds and Ulanta.
\end{IEEEbiography}

\begin{IEEEbiography}[{\includegraphics[width=1in,height=1.25in,clip,keepaspectratio]{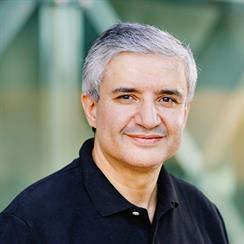}}]{MOHAMED-SLIM ALOUINI} (S'94-M'98-
	SM'03-F'09) was born in Tunis, Tunisia.
	He received the Ph.D. degree in electrical engineering from the California Institute of Technology (Caltech), Pasadena, CA, USA, in 1998.
	He has served as a Faculty Member with the
	University of Minnesota, Minneapolis, MN, USA,
	and with the Texas A\&M University at Qatar, Education City, Doha, Qatar, before joining the King
	Abdullah University of Science and Technology
	(KAUST), Thuwal, Saudi Arabia, as a Professor of electrical engineering,
	in 2009. His current research interests include the modeling, design, and
	performance analysis of wireless communication systems
\end{IEEEbiography}

\begin{IEEEbiography}[{\includegraphics[width=1in,height=1.25in,clip,keepaspectratio]{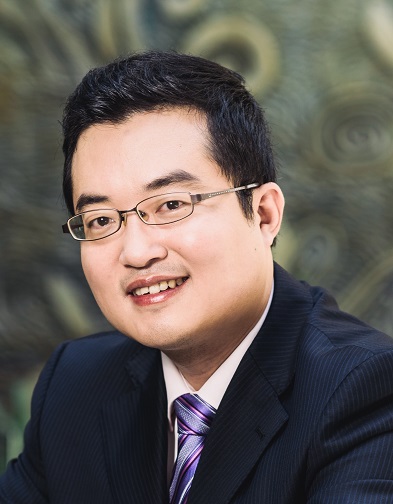}}]{Rui Zhang} Rui Zhang (S'00-M'07-SM'15-F'17) received the
	B.Eng. (first-class Hons.) and M.Eng. degrees from
	the National University of Singapore, Singapore,
	and the Ph.D. degree from the Stanford University,
	Stanford, CA, USA, all in electrical engineering.
	From 2007 to 2010, he worked as a Research
	Scientist with the Institute for Infocomm Research,
	ASTAR, Singapore. Since 2010, he has joined the
	Department of Electrical and Computer Engineering,
	National University of Singapore, where he is now
	a Dean’s Chair Associate Professor in the Faculty
	of Engineering. He has authored over 300 papers. He has been listed as a
	Highly Cited Researcher (also known as the World’s Most Influential Scientific
	Minds), by Thomson Reuters since 2015. His research interests include
	wireless information and power transfer, drone communication, wireless
	eavesdropping and spoofing, energy-efficient and energy-harvesting-enabled
	wireless communication, multiuser MIMO, cognitive radio, and optimization
	methods.
	
	He was the recipient of the 6th IEEE Communications Society Asia-Pacific
	Region Best Young Researcher Award in 2011, and the Young Researcher
	Award of National University of Singapore in 2015. He was the co-recipient
	of the IEEE Marconi Prize Paper Award in Wireless Communications in 2015,
	the IEEE Communications Society Asia-Pacific Region Best Paper Award in
	2016, the IEEE Signal Processing Society Best Paper Award in 2016, the
	IEEE Communications Society Heinrich Hertz Prize Paper Award in 2017,
	the IEEE Signal Processing Society Donald G. Fink Overview Paper Award
	in 2017, and the IEEE Technical Committee on Green Communications \&
	Computing (TCGCC) Best Journal Paper Award in 2017. His coauthored paper
	received the IEEE Signal Processing Society Young Author Best Paper Award
	in 2017. He served for over 30 international conferences as TPC Co-Chair or
	Organizing Committee Member, and as the guest editor for 3 special issues
	in IEEE Journal of Selected Topics in Signal Processing and IEEE Journal
	on Selected Areas in Communications. He was an elected member of the
	IEEE Signal Processing Society SPCOM (2012-2017) and SAM (2013-2015)
	Technical Committees, and served as the Vice Chair of the IEEE Communications Society Asia-Pacific Board Technical Affairs Committee (2014-2015).
	He served as an Editor for the \textit{IEEE Transactions on Wireless Communications} (2012-2016), the \textit{IEEE Journal on Selected Areas in Communications}: Green Communications and Networking
	Series (2015-2016), and the \textit{IEEE Transactions on Signal Processing} (2013-2017). He is now an Editor for the \textit{IEEE Transactions on Communications} and the \textit{IEEE Transactions on Green Communications and Networking}. He serves as a member of the
	Steering Committee of the \textit{IEEE Wireless Communications Letters}.
\end{IEEEbiography}

\end{document}